\documentclass[fleqn,usenatbib]{mnras}

\usepackage{newtxtext,newtxmath}
\usepackage[T1]{fontenc}
\usepackage{ae,aecompl}
\usepackage{pdflscape}
\usepackage{afterpage}

\usepackage{graphicx}	% Including figure files
\usepackage{amsmath}
\usepackage{amssymb}	% Extra maths symbols
\newcommand\target{{MAXI~J1348$-$630}}
\newcommand\nicer{{\it NICER}}

\defcitealias{Zhang2020a}{Paper-I}
\title[Transient type-B QPOs in \target]{NICER uncovers the transient nature of the type-B quasi-periodic oscillation in the black hole candidate \target}

\author[Zhang et al.]{L. Zhang$^{1},$\thanks{E-mail: liang.zhang@soton.ac.uk}
D. Altamirano$^{1}$,
P. Uttley$^{2}$,
F. Garc\'ia$^{3}$,
M. M\'{e}ndez$^{3}$,
J. Homan$^{4,5}$,
\newauthor
J. F. Steiner$^{6}$,
K. Alabarta$^{1,3}$,
D. J. K. Buisson$^{1}$,
R. A. Remillard$^{7}$,
\newauthor
K. C. Gendreau$^{8}$,
Z. Arzoumanian$^{8}$,
C. Markwardt$^{8}$,
T. E. Strohmayer$^{8}$,
\newauthor
J. Neilsen$^{9}$ 
and
A. Basak$^{2}$ \\
$^{1}$Physics and Astronomy, University of Southampton, Southampton, Hampshire SO17 1BJ, UK \\
$^{2}$Anton Pannekoek Institute for Astronomy, University of Amsterdam, Science Park 904, 1098 XH Amsterdam, The Netherlands\\
$^{3}$Kapteyn Astronomical Institute, University of Groningen, PO Box 800, NL-9700 AV Groningen, the Netherlands\\
$^{4}$Eureka Scientific, Inc., 2452 Delmer Street, Oakland, CA 94602, USA \\
$^{5}$SRON, Netherlands Institute for Space Research, Sorbonnelaan 2, 3584 CA Utrecht, The Netherlands \\
$^{6}$Harvard-Smithsonian Center for Astrophysics, 60 Garden St., Cambridge, MA 02138, USA \\
$^{7}$MIT Kavli Institute for Astrophysics and Space Research, MIT, 70 Vassar Street, Cambridge, MA 02139, USA \\
$^{8}$Astrophysics Science Division, NASA Goddard Space Flight Center, Greenbelt, MD 20771, USA \\
$^{9}$Villanova University, Villanova, PA 19085, USA \\
}

\date{Accepted 2021 May 25. Received 2021 May 24; in original form 2021 March 3}

\pubyear{2015}

\begin{document}
\label{firstpage}
\pagerange{\pageref{firstpage}--\pageref{lastpage}}
\maketitle

\begin{abstract}
We present a systematic spectral-timing analysis of a fast appearance/disappearance of a type-B quasi-periodic oscillation (QPO), observed in four \nicer\ observations of \target. By comparing the spectra of the period with and without the type-B QPO, we found that the main difference appears at energy bands above $\sim$2 keV, suggesting that the QPO emission is dominated by the hard Comptonised component. During the transition, a change in the relative contribution of the disk and Comptonised emission was observed. The disk flux decreased while the Comptonised flux increased from non-QPO to type-B QPO. However, the total flux did not change too much in the \nicer\ band. Our results reveal that the type-B QPO is associated with a redistribution of accretion power between the disk and Comptonised emission. When the type-B QPO appears, more accretion power is dissipated into the Comptonised region than in the disk. Our spectral fits give a hint that the increased Comptonised emission may come from an additional component that is related to the base of the jet. 
\end{abstract}


%%%%%%%%%%%%%%%%%%%%%%%%%%%%%%%%%%%%%%%%%%%%%%%%%%

%%%%%%%%%%%%%%%%% BODY OF PAPER %%%%%%%%%%%%%%%%%%

\section{Introduction}

Black hole binaries (BHBs), consisting of a stellar-mass black hole accreting material from a companion star, are typically transient systems. They spend most of their time in a long quiescent period ($L_{\rm X} \approx 10^{30}-10^{34}$ erg s$^{-1}$), and occasionally exhibit bright outbursts during which their X-ray luminosity increases by several orders of magnitude (typically reaching $L_{\rm X} \approx 10^{37}-10^{38}$ erg s$^{-1}$). 
BHBs in outbursts are characterized by rapid evolution of their X-ray spectral and timing properties, which can be studied in terms of different states (see \citealt{Homan2005,Remillard2006} for different classification schemes of black hole states). Two main canonical states are usually identified: the hard state (HS), where the spectrum is dominated by a hard Comptonised component, and the soft state (SS), where the X-ray emission is dominated by a thermal disk component. During the transition between the HS and SS, two additional short-lived states, namely the hard-intermediate state (HIMS) and the soft-intermediate state (SIMS), have been recognized (see \citealt{Belloni2016} for a recent review).

A distinctive timing characteristic of BHBs is the evolution of the low-frequency QPOs with spectral states (see \citealt{Ingram2019} for a recent review of different types of QPOs).
As a source evolves from the bright end of the HS to the HIMS, type-C QPOs (up to 20 per cent rms in the 2--60 keV, the full energy band of the Proportional Counter Array aboard the Rossi X-ray Timing Explorer, hereafter {\it RXTE}/PCA) always appear in the power density spectra (PDS), superimposed on a strong band-limited noise component. Their frequency is tightly correlated with spectral changes, moving from a few mHz to 10 Hz as the spectrum softens. Type-C QPOs are thought to have a geometric origin \citep[e.g.,][]{Motta2015,Eijnden2017,Ma2021}, and have been widely explained by Lense-Thirring precession of the hot inner flow \citep[e.g.,][]{Schnittman2006,Ingram2009}.
Type-B and type-A QPOs are normally detected in the SIMS, at a similar frequency range (typically 4--8 Hz). Both types are associated with weak broadband noise. Type-B QPOs are relatively strong (up to 5 per cent rms in the {\it RXTE}/PCA band) and narrow, whereas type-A QPOs are weak (few per cent rms in the {\it RXTE}/PCA band) and broad features. The physical origin of these two types of QPOs is less clear.
Hints of a connection between the appearance of type-B QPO and the launch of discrete jet ejection have been found in several BHBs \citep[e.g.,][]{Soleri2008,Fender2009,Russell2019A,Homan2020,Russell2020}, pointing to a jet-based origin of type-B QPOs.
In this work, we will focus our study on the type-B QPOs.

QPOs are typically transient, appearing or disappearing on timescales of a few tens of seconds. Such transient behaviour has been observed in all type-A \citep{Sriram2021}, -B \citep[e.g.,][]{Miyamoto1991,Nespoli2003,Casella2004,Huang2018,Sriram2021}, and -C \citep{Xu2019,Jithesh2019,Sriram2021} QPOs.
In particular, fast transitions between different types of low-frequency QPOs or broadband noise components are sometimes observed with significant changes of the source spectrum \citep[e.g.,][]{Miyamoto1991,Takizawa1997,Wijnands2001,Homan2001,Casella2004,Motta2011,Sriram2012,Sriram2013,Sriram2016,Sriram2021}.
The spectral-timing study of these fast transitions not only provides important evidence on the physical origin of different types of QPOs, but also helps us better understand the mechanism responsible for the state transitions.

Fast QPO transitions were first discovered in {\it Ginga} data, in GX 339$-$4 \citep{Miyamoto1991} and GS 1124$-$68 \citep{Takizawa1997} in those observations showing ``flip-flop" variations. A 6-Hz type-B QPO, which was visible when the source flux is high, disappeared when the source moved to a low-flux epoch. The transition from the low-flux to high-flux epoch was mainly due to a change in the Comptonised component \citep{Miyamoto1991}. 
Later with {\it RXTE}, fast transitions between type-B QPOs and type-A QPOs have been observed in a few BHBs, e.g, GX 339$-$4 \citep{Nespoli2003,Motta2011}, XTE J1817$-$330 \citep{Sriram2012}, XTE J1550$-$564 \citep{Homan2001,Sriram2016} and XTE J1859+226 \citep{Casella2004,Sriram2013}. 
Type-C to type-B transitions or vice versa are more frequently seen when the source evolves from the HIMS to the SIMS \citep[e.g.,][]{Motta2011,Homan2020,Sriram2021} or from the SIMS back to the HIMS. 
Direct transitions between type-C QPOs and type-A QPOs were recently found in Swift J1658.2--4242 during the flip-flop variations \citep{Bogensberger2020}.

\begin{figure}
    \begin{center}
        \resizebox{\columnwidth}{!}{\rotatebox{0}{\includegraphics[clip]{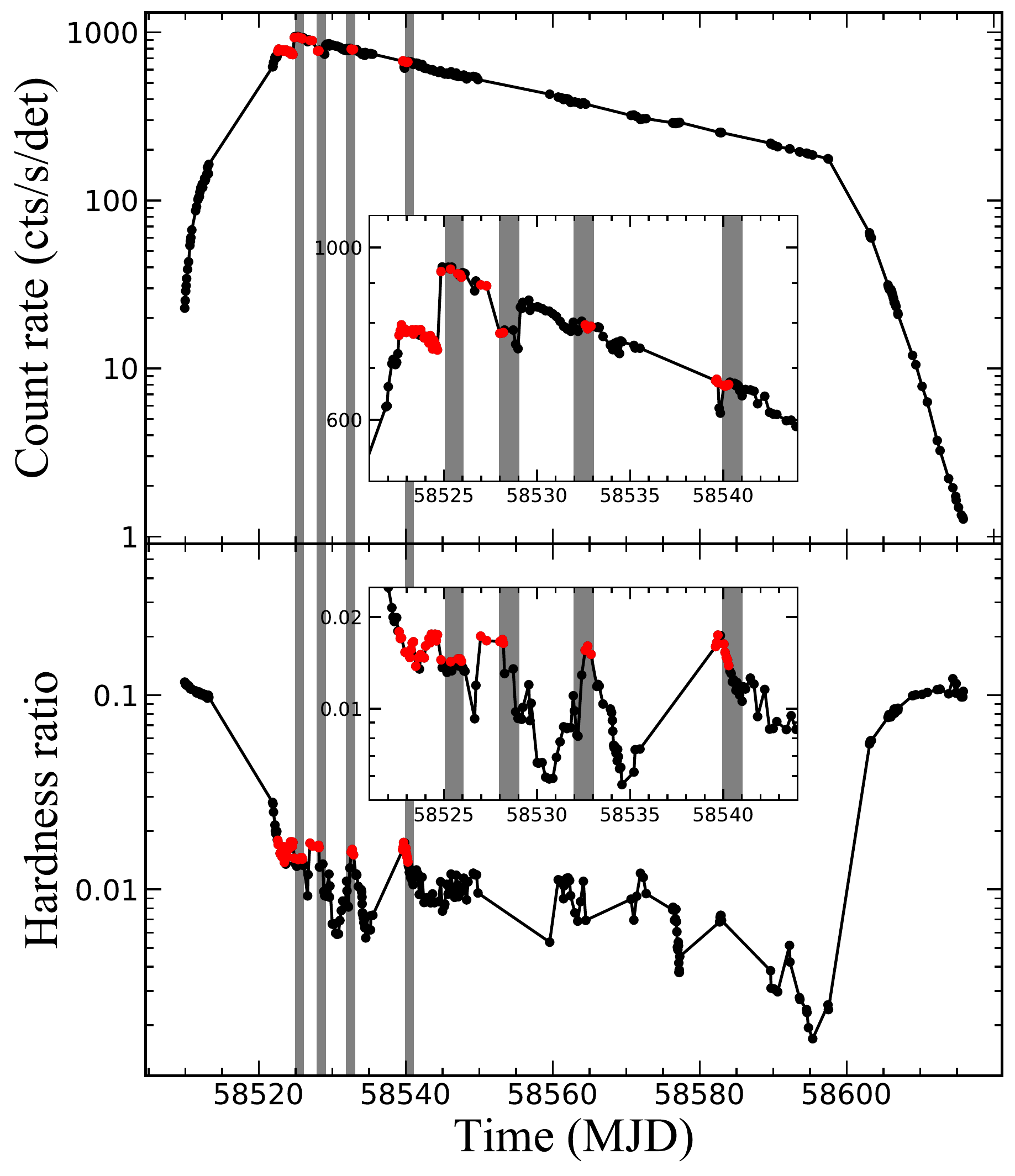}}}
    \end{center}
    \caption{Long-term evolution of the main outburst of \target. Upper panel: \nicer\ 0.5--12 keV light curve. Lower panel: time evolution of the hardness ratio (6--12 keV/2--3.5 keV). Each data point represents the average of a single ISS-orbit dwell. The light curves were normalised by the number of active detectors. The red points correspond to the orbits where a type-B QPO is present. The grey bands mark the four observations in which a fast appearance/disappearance of a type-B QPO was observed. The inset panels show a magnification of the region containing the fast transitions.}
    \label{fig:long_term_lc}
\end{figure}

\target\ is a bright (above 3 Crab at the outburst peak) X-ray transient discovered by {\it MAXI}/GSC on 2019 January 26 \citep{Yatabe2019,Tominaga2020}. 
Following observations of {\it Swift}/XRT obtained a precise source position \citep{Kennea2019} and confirmed the optical counterpart identified by \citet{Denisenko2019}. Observations of the Australia Telescope Compact Array (ATCA) detected a radio counterpart consistent with the X-ray position \citep{Russell2019B}.
Based on a spectral-timing analysis of observations from the {\it Neutron Star Interior Composition Explorer} \citep[{\it NICER},][]{Gendreau2016}, \citet[hereafter \citetalias{Zhang2020a}]{Zhang2020a} concluded that \target\ is a  black hole candidate in a binary system. 
The source underwent an initial full outburst which showed all canonical spectral states of BHBs. After the main outburst, it re-brightened a few times while staying in the HS \citep{Negoro2019,Zhang2020a,Zhang2020b,Tominaga2020}.
\citet{Tominaga2020} reported the outburst evolution of \target\ as observed with {\it MAXI}. They noted that the source is in the direction of the Galactic Scutum-Centaurus arm, and due to the low column density it may lie in front of the arm at about 3--4 kpc.
\citet{Chauhan2020} determined the distance of \target\ to be $2.2_{-0.6}^{+0.5}$ kpc using H\,{\sc i} absorption.
\citet{Carotenuto2021} reported the radio monitoring of \target\  with MeerKAT and ATCA. They detected the rise, quenching, and re-activation of the compact jet, as well as two single-sided discrete ejections during the main outburst.
Different types of low-frequency QPOs have been observed at different phases of the outburst (see \citetalias{Zhang2020a} for details). 
Type-B QPOs were detected in the SIMS at a stable frequency of $\sim$4.5 Hz. \citet{Belloni2020} studied the energy dependence of the fractional rms and phase lags of the type-B QPO using \nicer\ data, and \citet{Garcia2021} successfully fitted them simultaneously with a two-component Comptonisation model.
\citet{Jithesh2021} studied the energy dependence of the fractional rms of a type-C QPO using simultaneous {\it AstroSat} and \nicer\ observations. They found that the rms of the QPO fundamental is nearly energy independent, while the rms of the subharmonic decreases with energy.

In this work, we present a systematic spectral-timing analysis of the fast appearance/disappearance of a type-B QPO, observed in four \nicer\ observations of \target. The main purpose of this work is to investigate the transient nature and the physical origin of the type-B QPO.

\section{Observations and Data Analysis}

\begin{table}
	\centering
	\caption{Log of the four \nicer\ observations in which a fast appearance/disappearance of a type-B QPO was observed. The last column lists the exposure time after data filtering and cleaning. The number in parentheses lists the exposure time of the period with a type-B QPO.}
	\label{tab:log}
	\begin{tabular}{lcccc} % four columns, alignment for each
		\hline\hline
		Case & ObsId & Start time & Exposure (s)\\
		\hline
	    \#1	& 1200530110 & 2019 Feb 11 02:17:13 & 6916 (2944)\\
		\#2 & 1200530113 & 2019 Feb 13 23:54:20 & 3089 (601)\\
		\#3 & 1200530117 & 2019 Feb 18 00:08:00 & 3558 (1956)\\
		\#4 & 1200530125 & 2019 Feb 26 00:41:12 & 11421 (2715)\\
		\hline
	\end{tabular}
\end{table}

\begin{figure*}
    \begin{center}
        \resizebox{2\columnwidth}{!}{\rotatebox{0}{\includegraphics[clip]{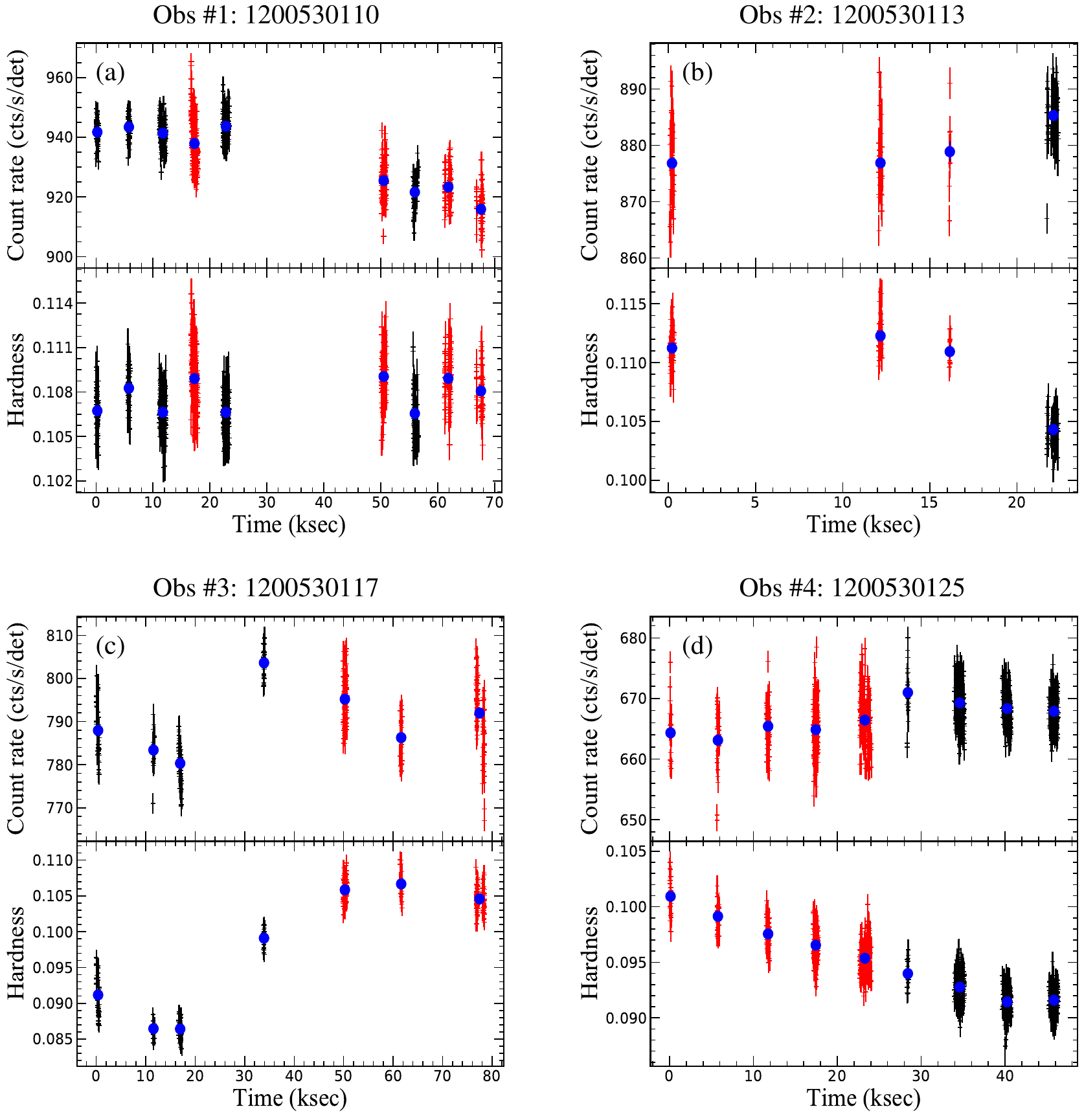}}}
    \end{center}
    \caption{\nicer\ 0.5--12 keV light curve and hardness ratio (3--12 keV/0.5--3 keV) with 5-s binning for the four observations in which a fast transition of a type-B QPO was observed. The light curves were normalised by the number of active detectors. The red (black) points represent the orbits with (without) a type-B QPO. The blue points are the average of a single orbit.}
    \label{fig:lc}
\end{figure*}

\begin{figure*}
    \begin{center}
        \resizebox{2\columnwidth}{!}{\rotatebox{0}{\includegraphics[clip]{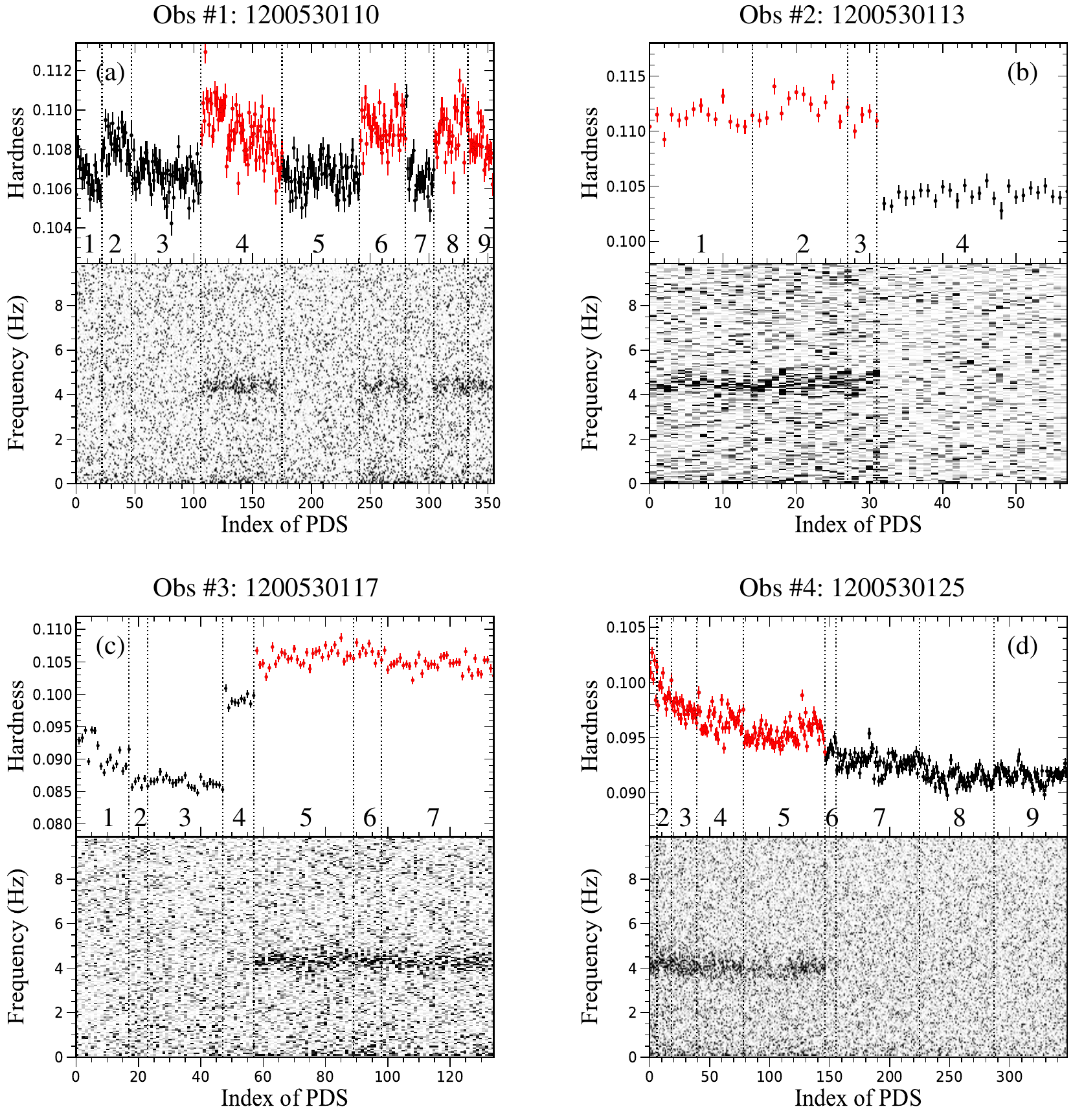}}}
    \end{center}
    \caption{\nicer\ hardness ratio (3--12 keV/0.5--3 keV) and corresponding dynamical power spectra with a time resolution of 16 s. All time gaps were removed and marked with dotted lines. The $x$-axis represents the index of each 16-s PDS. The red (black) points represent the period with (without) a type-B QPO. For the dynamical power spectra, the powers were not rms-normalised and Poisson noise was not subtracted. Darker shades correspond to stronger variability. }
    \label{fig:dpds}
\end{figure*}

\begin{figure*}
    \begin{center}
        \resizebox{2\columnwidth}{!}{\rotatebox{0}{\includegraphics[clip]{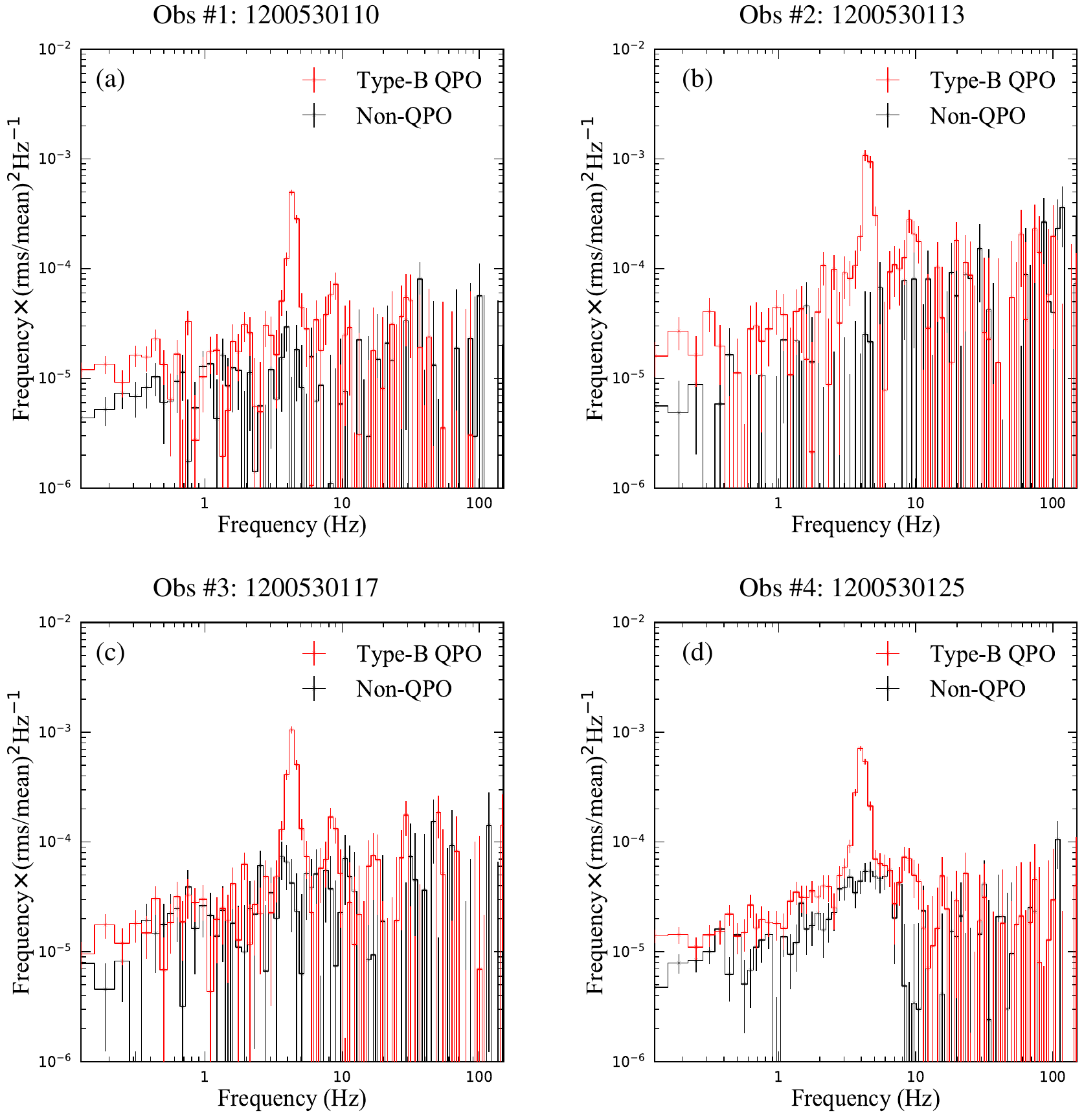}}}
    \end{center}
    \caption{Power spectra averaged from the period with (red) and without (black) a type-B QPO. The average PDS were calculated in the 0.5--12 keV energy band, and rms-normalised \citep{Belloni1990}. The contribution due to Poisson noise was subtracted. }
    \label{fig:pds}
\end{figure*}

\nicer\ is a soft X-ray telescope onboard the International Space Station (ISS) launched in 2017 June. It provides high throughput in the 0.2--12 keV energy band with an absolute timing precision of $\sim$100 ns, making it an ideal instrument to study fast X-ray variability. \nicer\ consists of 56 co-aligned concentrator X-ray optics, each paired with a single-pixel silicon drift detector. Presently, 52 detectors are active with a peak effective area of $\sim$1900 ${\rm cm}^2$ at 1.5 keV. In our analysis, we excluded data from detectors \#14 and \#34 as they occasionally show episodes of increased instrumental noise.

\nicer\ has extensively monitored the outburst of \target\ in great detail since its discovery. 
%
%We used all the observations taken during the initial full outburst, between 2019 January 26 and 2019 May 21.
%
All observations were reprocessed using the \nicer\ software tools {\sc nicerdas} version 6.0, distributed with {\sc heasoft} version 6.26. The data reduction procedure was the same as described in \citetalias{Zhang2020a}.
To follow the outburst evolution (excluding reflares, see fig. 1 in \citetalias{Zhang2020a}), we extracted net light curves in the 0.5--12 keV energy band as well as the 2--3.5 keV and 6--12 keV energy bands using {\sc xselect}. In the upper panel of Fig. \ref{fig:long_term_lc} we show the 0.5--12 keV light curve, and in the lower panel we show the evolution of the hardness ratio (6--12 keV/2--3.5 keV) during the main outburst.
Individual \nicer\ observations are subdivided by continuous data segments, typically separated by several ksec gaps due to the orbit of the ISS. In Fig. \ref{fig:long_term_lc}, each data point represents the average from a single ISS orbit (we define an orbit as a continuous data segment with a net exposure time of longer than 60 s of good time).
During the brightest phase of the outburst, some of the detectors were switched off to prevent telemetry saturation and accommodate the high source flux. Therefore, we have normalised the intensity by the number of active detectors in all figures of this paper.

In \citetalias{Zhang2020a}, we produced a PDS in the 0.5--12 keV energy band for each orbit. After the source entered the SIMS, a significant type-B QPO was observed in a number of orbits (see \citetalias{Zhang2020a} and \citealt{Belloni2020} for details). In Fig. \ref{fig:long_term_lc}, the orbits containing a type-B QPO are marked in red. 
The type-B QPO appears in a narrow hardness range, and always corresponds to local peaks in hardness ratio. Upon inspection, we found that the type-B QPOs are not present all the time during the SIMS, and they sometimes switch on/off between orbits. Table \ref{tab:log} lists the log of the four observations in which we found a fast transition. 
The time intervals of these transitions are marked with grey shaded areas in Fig. \ref{fig:long_term_lc}. In the remaining of the paper, we will focus our analysis on the characteristics of these four observations.

\subsection{Timing and spectral analysis}

For each of the four observations, we produced a light curve with 5-s binning in the 0.5--12 keV energy band. To calculate the hardness ratio, we also extracted light curves in the 3--12 keV and 0.5--3 keV energy bands. 
The light curves were not background-subtracted since the background contribution (about 2 counts s$^{-1}$) is negligible during the brightest phase of the outburst.

PDS shown in this paper were calculated in the 0.5--12 keV energy band with 16-s long intervals and 1/8192-s time resolution ($16^{-1}-4096$ Hz frequency range). The resulting PDS were averaged and then normalized in units of ${\rm {(rms/mean)}^2~Hz^{-1}}$ \citep{Belloni1990}. The contribution due to Poisson noise was subtracted. We fitted the PDS with a model consisting of a sum of Lorentzian functions.

For the spectral analysis, we extracted background-subtracted spectra using the \nicer\ Background Estimator Tool ``nibackgen3C50"\footnote{\url{https://heasarc.gsfc.nasa.gov/docs/nicer/tools/nicer_bkg_est_tools.html}}. The spectra were corrected for the oversampling and rebinned to have at least 30 counts per bin. As suggested by the \nicer\ team (private communication), we added a systematic error of 3\% below 1.5 keV and 0.5\% above 1.5 keV using {\sc grppha}. We fitted the spectra in the 0.6--10 keV band using {\sc xspec} version 12.10.1.
We used the latest calibration files obtained from the {\sc caldb} version 20200722, downloaded from NASA's High Energy Astrophysics Science Archive Research Center (HEASARC).

\begin{table*}
    \centering
    \caption{Properties of the type-B QPOs for the four observations in which a fast appearance/disappearance of a type-B QPO was observed. The fractional rms of the QPO was calculated in the 0.5--12 keV energy band. The second and third columns list the total count rate and the number of active detectors, respectively.}
    \label{tab:qpo_para}
    \begin{tabular}{lccccccccc}
        \hline\hline
        ObsID      &   Total Count Rate  &  Num\_Det\_On   &   \multicolumn{3}{c}{QPO Fundamental}   &&   \multicolumn{3}{c}{Second Harmonic} \\
        \cline{4-6} \cline{8-10} 
                   &  (counts s$^{-1}$)  &                 &   $\nu_{0}$ (Hz) &   $Q$   &  rms (\%)  &&  $\nu_{0}$ (Hz) &   $Q$   &  rms (\%) \\
        \hline
        1200530110  &        24157       &   26            & $4.37 \pm 0.03$   & $7.93 \pm 0.51$ & $0.98 \pm 0.03$ 
                                                           && $8.54 \pm 0.32$  & $5.15 \pm 4.03$ & $0.40 \pm 0.07$ \\
        \hline
        1200530113  &        20177       &   23            & $4.48 \pm 0.04$   & $8.13 \pm 1.44$ & $1.59 \pm 0.06$ 
                                                           && $8.89 \pm 0.18$  & $5.89 \pm 2.53$ & $0.89 \pm 0.02$ \\
        \hline
        1200530117  &        20608       &   26            & $4.30 \pm 0.02$   & $8.14 \pm 0.86$ & $1.48 \pm 0.04$ 
                                                           && $8.57 \pm 0.23$  & $8.07 \pm 5.60$ & $0.58 \pm 0.08$ \\
        \hline
        1200530125  &        27283       &   41            & $4.06 \pm 0.01$   & $6.40 \pm 0.36$ & $1.37 \pm 0.02$ 
                                                           && $8.57 \pm 0.25$  & $3.90 \pm 1.65$ & $0.52 \pm 0.07$ \\
        \hline
    \end{tabular}
\end{table*}

\section{Results}

\subsection{Detailed look at the fast transitions}

\subsubsection{Light curve and hardness ratio}

In Fig. \ref{fig:lc}, we show the 0.5--12 keV light curve and the evolution of the hardness ratio (3--12 keV/0.5--3 keV) with 5-s binning for the four observations in which a fast appearance/disappearance of a type-B QPO was observed. 
Obs. \#2 contains a total of 7 orbits. We only show the results of the first four orbits in this work, as orbit 5 was taken near half day later than orbit 4 and no QPOs were detected in the remaining three orbits. During the last three orbits, the average count rate decreased from $884 \pm 7$ to $836 \pm 7$ counts s$^{-1}$ det$^{-1}$. The hardness ratio decreased from 0.105 to 0.090.
In the case of Obs. \#4, a total of 15 orbits were performed in this observation. We ignored the last six orbits without any QPOs. The average count rate decreased slightly from $664 \pm 5$ to $656 \pm 5$ counts s$^{-1}$ det$^{-1}$ during the last six orbits. The hardness ratio remained at around 0.090. 
%
% In Obs. \#1, the source count rate (upper panel of Fig. \ref{fig:lc}a) remained approximately constant at $\sim$940 cts/s/det in the first five orbits, while it was slightly lower at $\sim$920 cts/s/det in the last four orbits. The hardness ratio (lower panel of Fig. \ref{fig:lc}a) did not change appreciably during the whole observation. 
% %
% During the period we show here, the source count rate (upper panel of Fig. \ref{fig:lc}b) remained more or less constant at $\sim$880 cts/s/det. 
% %
% The hardness ratio (lower panel of Fig. \ref{fig:lc}b) remained at $\sim$0.111 during the first three orbits, while it showed a clear decrease to $\sim$0.104 at orbit 4.
% %
% The behaviour of the light curve and hardness ratio in Obs. \#3 is complex. During the first three orbits, the source count rate (upper panel of Fig. \ref{fig:lc}c) decreased slightly with a minor drop in hardness ratio (lower panel of Fig. \ref{fig:lc}c). After that, a clear increase in source count rate was seen at orbit 4 with spectral hardening. After orbit 4, the source count rate showed a slight decrease again until the last orbit where it was a little higher than the previous orbit. The hardness ratio kept increasing and reached its maximum at orbit 5, after which the hardness ratio remained approximately constant at $\sim$0.106. 
% %
% During the period we show, the source count rate (upper panel of Fig. \ref{fig:lc}d) stayed approximately constant at $\sim665$ cts/s/det. The hardness ratio (lower panel of Fig. \ref{fig:lc}d) dropped gradually from $\sim$0.101 to $\sim$0.091.
%
As we show in section \ref{sec:dpds} below, the type-B QPOs were only detected in certain orbits. In Fig. \ref{fig:lc} we mark the orbits with the QPO with red, while the ones without the QPO with black.
We found the following: 
\begin{itemize}
    \item the turn-on of the type-B QPO does not appear to depend on the 0.5--12 keV count rate as we found QPOs in orbits at higher and lower averaged count rate than those without a significant QPO (see, e.g., Obs \#1 and \#3);
    \item within each observation, the orbits with the type-B QPO always correspond to a higher hardness ratio than those without the QPO;
    \item the hardness value corresponding to the transition changes between observations, possibly decreasing with decreasing count rate.
\end{itemize}

\begin{figure}
\begin{center}
\resizebox{\columnwidth}{!}{\rotatebox{0}{\includegraphics[clip]{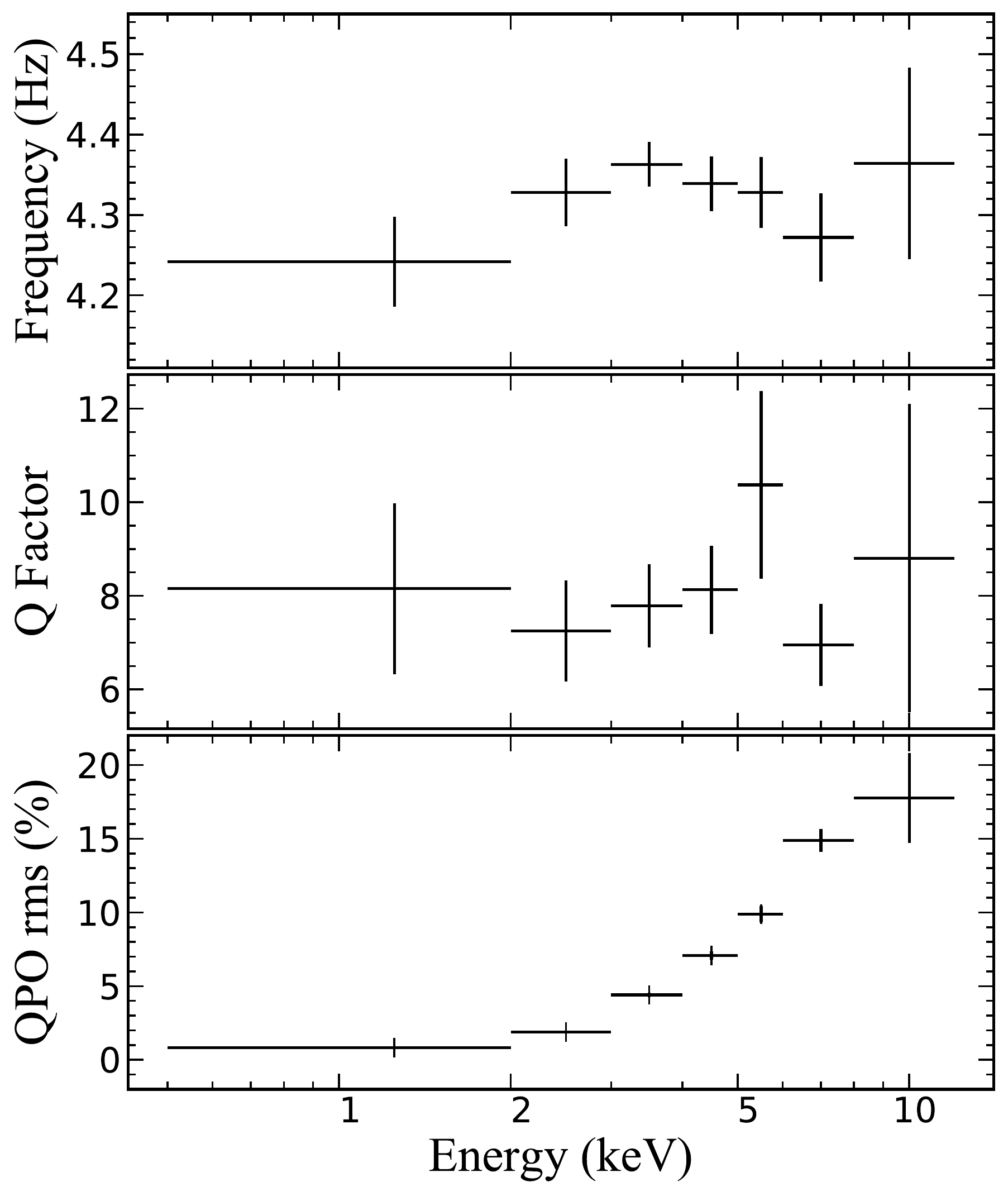}}}
\end{center}
\caption{Energy dependence of the frequency, $Q$ factor, and fractional rms of the type-B QPO observed in Obs \#3: 1200530117.}
\label{fig:energy_dependence}
\end{figure}

\begin{figure*}
\begin{center}
\resizebox{2\columnwidth}{!}{\rotatebox{0}{\includegraphics[clip]{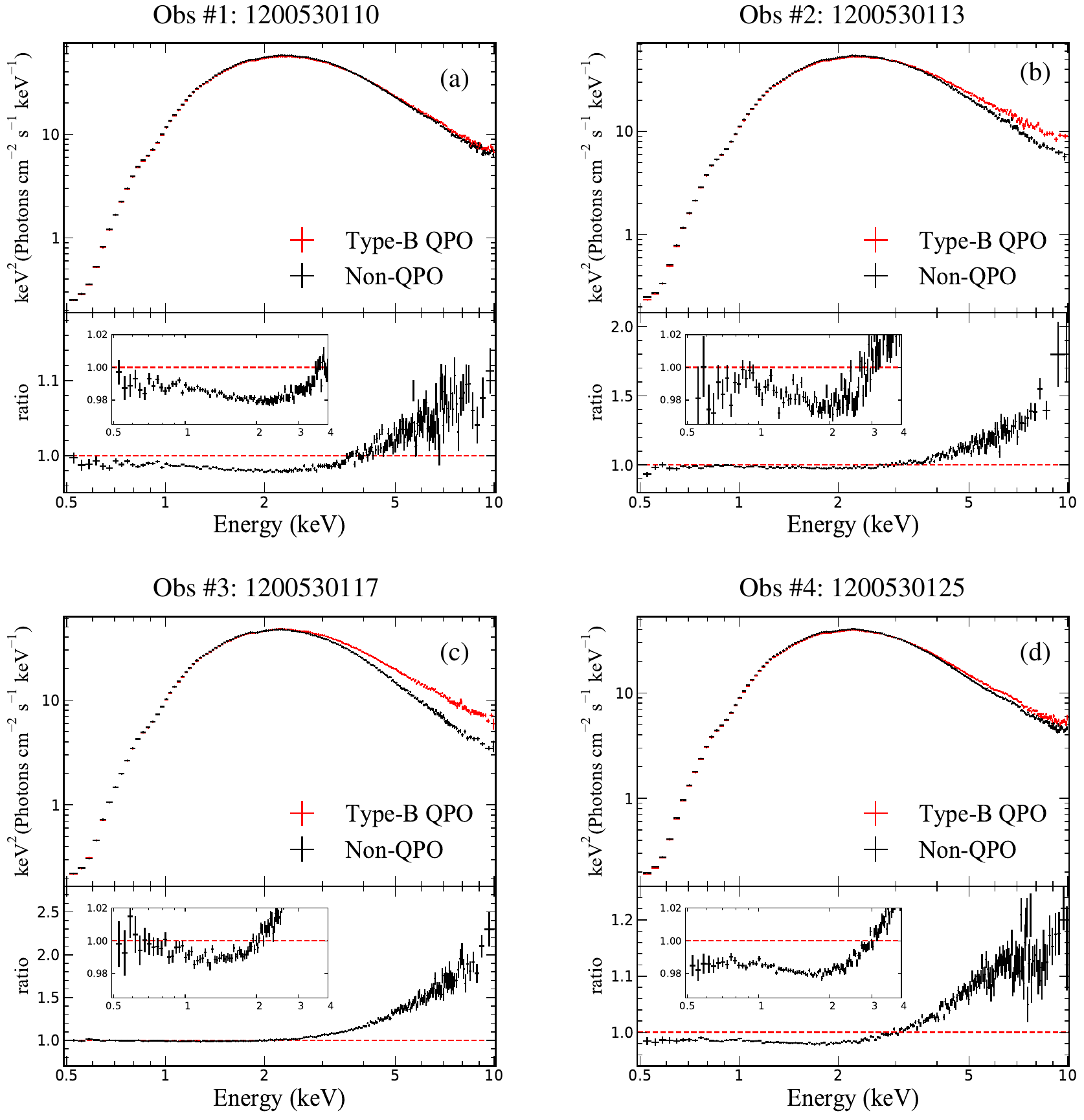}}}
\end{center}
\caption{Unfolded spectra of the period with (red) and without (black) a type-B QPO, and corresponding spectral ratio defined as the ratio between the spectrum with the type-B QPO and that without the QPO (type-B/non-QPO). The spectra were extracted from all the orbits with and without the type-B QPO, respectively, and deconvolved against a power law with $\Gamma = 2$.}
\label{fig:spectra_average}
\end{figure*}

\begin{figure*}
\begin{center}
\resizebox{2\columnwidth}{!}{\rotatebox{0}{\includegraphics[clip]{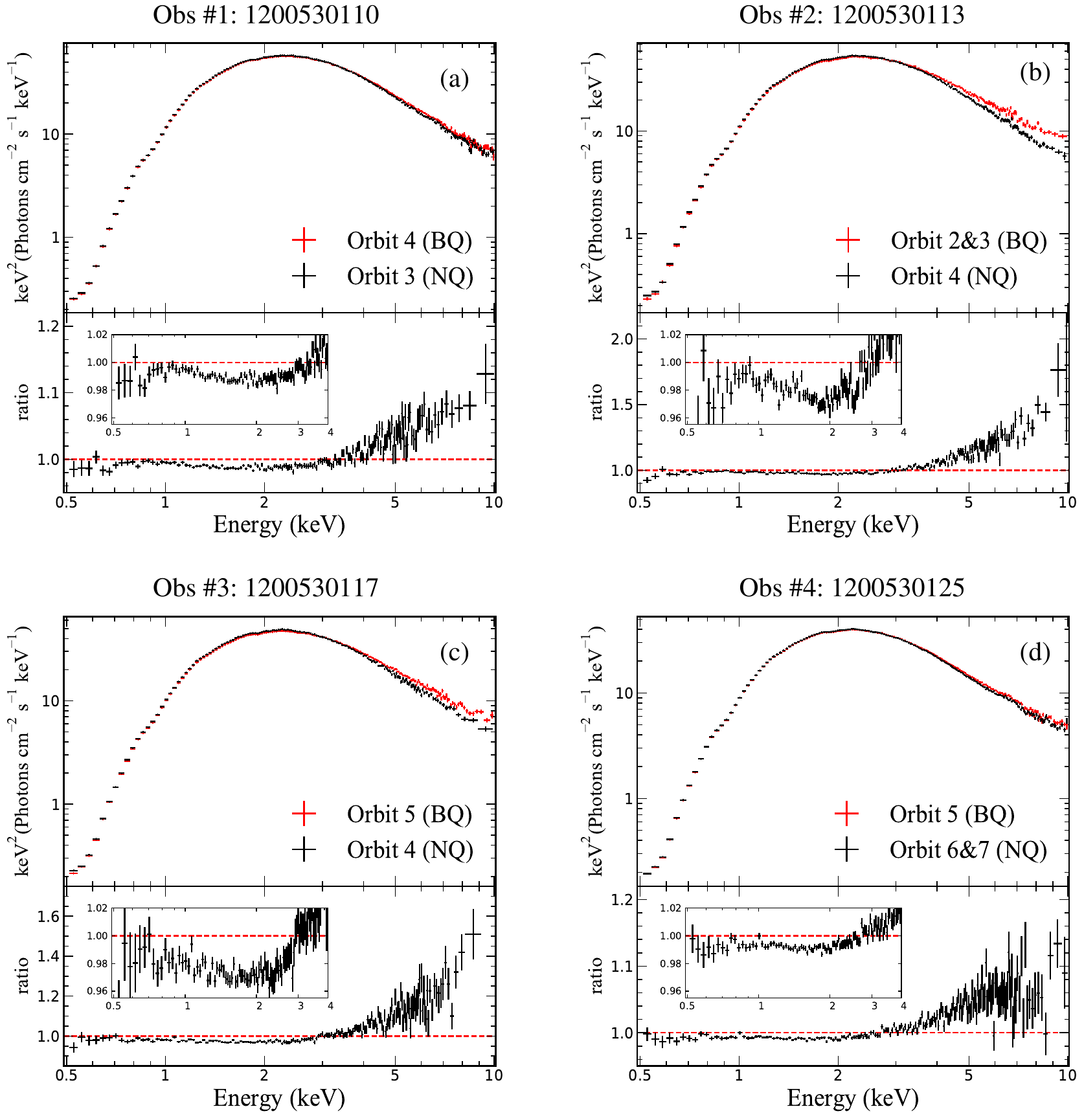}}}
\end{center}
\caption{Same as Fig. \ref{fig:spectra_average}. However, the spectra of the period with and without the type-B QPO were extracted from one or two single orbits before and after the transition.}
\label{fig:spectra_orbit}
\end{figure*}

\subsubsection{Dynamical and average power spectra}\label{sec:dpds}

In Fig. \ref{fig:dpds} we show the 0.5--12 keV light curve and corresponding dynamical PDS (with a time resolution of 16 s) for each of the four observations shown in Fig. \ref{fig:lc}. 
All time gaps were removed and marked with dotted lines. The $x$-axis label is the index of each 16-s PDS.
The dynamical PDS reveals that a significant transient QPO around 4 Hz was detected in each observation.
In Obs. \#1, the QPO was visible in orbits 4, 6, 8 and 9, and disappeared in the other orbits. In Obs. \#2 and \#4, the QPO was only seen in the first three and five orbits, respectively. While in Obs. \#3, the QPO was undetectable in the first four orbits, but appeared in the last three orbits.
In all cases, the frequency of the QPO remained more or less constant within each observation.
%
% We did not find fast transitions within an orbit. This is not surprising, as the transition from non-QPO to a type-B one can be as fast as a few tens of seconds \citep{Nespoli2003}.

Since the dynamical PDS shows that the QPO frequency is stable, we made an average PDS per observation from all the orbits with and without the QPO, respectively.  The resulting PDS are shown in Fig. \ref{fig:pds} and the best-fitting parameters of the QPO are listed in Table \ref{tab:qpo_para}. Hereafter we refer to the period with and without the QPO as the BQ (type-B QPO) and NQ (non-QPO) epoch for convenience.
The average PDS of epoch BQ shows a primary peak at around 4.3 Hz and its harmonic at around 8.6 Hz. 
The fundamental of the QPO is very sharp ($Q \approx 6-8$) with a fractional rms of $\lesssim$1.6\% in the 0.5--12 keV energy band, and $\sim$4--6\% in the 3--12 keV energy band. The value of the 3--12 keV fractional rms is consistent with the typical value measured with {\it RXTE}/PCA at a similar energy band \citep[see, e.g.,][]{Casella2004,Motta2016}.
The lower fractional rms amplitude in the 0.5--12 keV energy band is likely due to dilution of the amplitude caused by the non-modulated photons from the disk component.
The second-harmonic peak was weak but significantly (>3$\sigma)$ detected in the average PDS.  
The frequency of the QPO and its second harmonic does not change much between observations (see Table \ref{tab:qpo_para}).
In addition to the QPO, we detected weak broadband noise with an integrated fractional rms of $\sim$1--2\% in the 0.1--20 Hz frequency range (0.5--12 keV).
Based on the properties of the QPO and noise, we can easily identify it as a type-B one \citep[see, e.g.,][and references therein]{Motta2016}. 
We also investigated the energy dependence of the QPO properties. In Fig. \ref{fig:energy_dependence} we show the frequency, $Q$ factor, and fractional rms of the QPO observed in Obs. \#3, as a function of photon energy. The QPO properties are similar in the other three cases. The frequency and $Q$ factor of the QPO do not change significantly with photon energy. The energy dependence of the fractional rms of the QPO is consistent with \citet{Belloni2020}: the QPO fractional rms increases from $\sim$1\% at below 2 keV to $\sim$20\% at around 10 keV.

The average PDS of epoch NQ are dominated by weak ($\lesssim$1\% in the 0.1--20 Hz frequency range) broadband noise without showing any narrow peak, similar to that usually observed in the SS of BHBs. 
To test the possibility of the presence of a QPO, we fitted the average PDS of epoch NQ using a Lorenzian function with the characteristic frequency and $Q$ factor fixed at the values of the type-B QPO detected in that observation. We measured a 3-$\sigma$ upper limit on the fractional rms of 0.31\%, 0.47\%, 0.55\% and 0.52\% for the four observations, respectively.

\subsubsection{Spectral difference}

To check the possible spectral changes during the transition, we extracted spectra from the period with and without the type-B QPO separately. Hereafter we refer to them as the BQ spectrum and the NQ spectrum, respectively.
In order to take advantage of the large amount of photons detected by \nicer, we first extracted spectra accumulated from the orbits with and without the QPO, for each observation. The BQ (red) and NQ (black) spectra and corresponding spectral ratio (BQ/NQ) are shown in Fig. \ref{fig:spectra_average}. 
This figure shows that even if the shape of the BQ (or NQ) spectrum is different between observations (also see the spectral fits in section \ref{sec:spectra_fit}), the shape of the spectral ratio is remarkably similar in all cases. 
The main difference appears at energy bands above 2 keV (but note that the ratio is not constant below 2 keV), where the spectral ratio increases gradually towards higher energies. This indicates the presence of a stronger hard component when the type-B QPO was detected. The value of the spectral ratio is much larger in Obs \#2 and \#3 (above 1.5 at 10 keV) than that in Obs \#1 and \#4 (around 1.15 at 10 keV). 
Below 2 keV, the spectral difference is subtle. Upon inspection, we found a slightly increasing trend towards lower energies, which is likely associated with variations in the disk component. The overall spectral ratio below 2 keV is smaller than unity, suggesting that the disk component is weaker during the appearance of the type-B QPO.
In Obs. \#1, the type-B QPO disappeared in the first three orbits, and reappeared in orbit 4, and then disappeared again in orbit 5. We compared the spectra of orbits 3 and 5. The spectra before the appearance of the type-B QPO and after its disappearance are the same within errors.

We note that the hardness ratio during the period with/without the QPO changes significantly in some observations (see Fig. \ref{fig:lc}), indicating spectral changes. 
To reduce the effect of these spectral changes, we also compared the BQ and NQ spectra by extracting the energy spectra from a single orbit (or two if its net exposure time is less than 150 s) before and after the transition. The unfolded spectra and corresponding spectral ratio are shown in Fig. \ref{fig:spectra_orbit}. The shape of the spectral ratio is consistent with that obtained from the average spectra.

\subsection{Spectral modelling}
\label{sec:spectra_fit}

\subsubsection{Fit with standard two-component continuum models}

In order to further study the spectral difference, we performed spectral modelling for the BQ and NQ spectra taking from one or two single orbits before and after the transition shown in Fig. \ref{fig:spectra_orbit}. For each observation, we fitted the BQ and NQ spectra jointly using two different models: 

\vspace{0.25cm}
\textbf{Model 1:}  \textsc{tbnew*(simpl*diskbb+gaussian)}, and   
\textbf{Model 2:}  \textsc{tbnew*(diskbb+nthcomp+gaussian)}. 
\vspace{0.25cm}

In both models, the absorption along the line of sight was modelled using {\sc tbnew}\footnote{\url{https://pulsar.sternwarte.uni-erlangen.de/wilms/research/tbabs/}}, which is a high resolution version of the X-ray absorption model {\sc tbabs}. All abundances were fixed at the solar value given by \citet{Wilms2000}\footnote{Note that in \citetalias{Zhang2020a}, the abundances we used are from \citet{Anders1989}. As a result, the $N_{\rm H}$ is lower than that measured in this work.}. 
The {\sc diskbb} component is a multi-temperature blackbody model accounting for emission from a geometrically thin, optically thick accretion disk \citep{Shakura1973}.
Both {\sc simpl} \citep{Steiner2009} and {\sc nthcomp} \citep{Zdziarski1996} describe the power-law like continuum produced by thermal Comptonisation of soft disk photons in a hot gas of electrons. {\sc simpl} is developed as a convolution model with two free parameters, i.e., the photon index, $\Gamma$, and the scattered fraction, $f_{\rm sc}$; whereas {\sc nthcomp} is used as an additive model. 
In model 2, we assumed that all the seed photons for the Comptonised component are produced by the disk and linked the seed photon temperature of the {\sc nthcomp} component, $kT_{\rm bb}$, to the inner disk temperature of the {\sc diskbb} component, $kT_{\rm in}$. Since the high-energy cutoff of the Comptonised component is outside the \nicer\ energy range, we fixed the electron temperature at 1000 keV. Therefore, the free parameters of the {\sc nthcomp} component are the photon index ($\Gamma$) and the normalisation ($N_{\rm nthcomp}$).
In both models, a {\sc gaussian} line was added to account for the weak Fe K$\alpha$ emission line seen in the spectra. In some cases, the line energy or width was fixed since the parameters could not be constrained very well. 
We allowed all the parameters to vary between the two epochs except for the equivalent hydrogen column density, $N_{\rm H}$, which was kept linked.
We calculated the 0.6--10 keV unabsorbed fluxes of different components using {\sc cflux} from model 2. 
The best-fitting spectral parameters for models 1 and 2 are listed in Table \ref{tab:best_fits_orbit}. Both models give acceptable spectral fits with $\chi^{2}/{\rm dof}<1.1$. In Fig. \ref{fig:corner_simpl} and \ref{fig:corner_nthcomp}, we show an example of the distributions of the main best-fitting spectral parameters for models 1 and 2, respectively.
For completeness, we also fitted the spectra taken from all the orbits with/without the QPO (shown in Fig. \ref{fig:spectra_average}) with models 1 and 2. We show the best-fitting results in Table \ref{tab:best_fits_avg}. The main conclusions of this paper are the same if we either fit the average or the single-orbit spectra.

\begin{figure}
\begin{center}
\resizebox{\columnwidth}{!}{\rotatebox{0}{\includegraphics[clip]{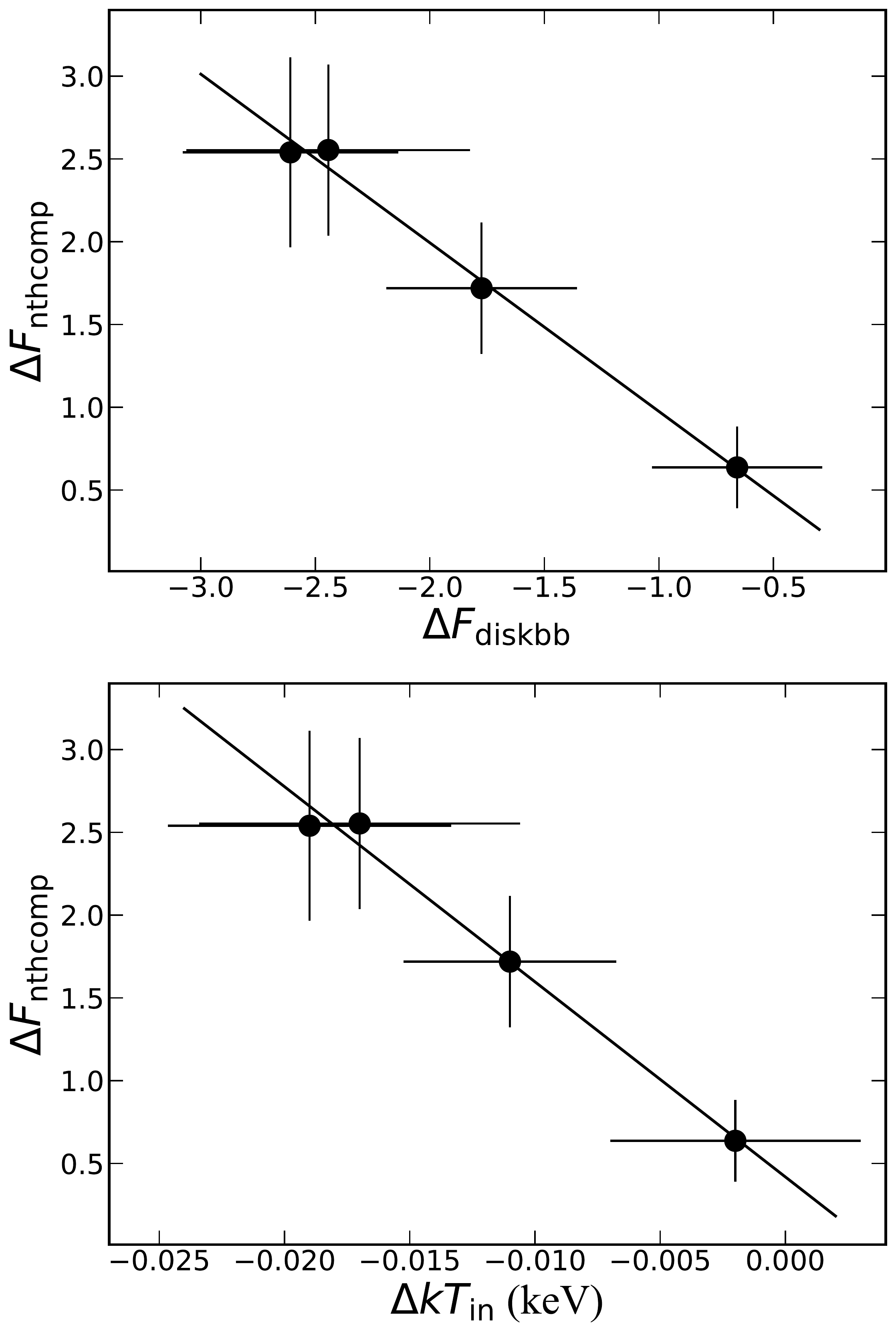}}}
\end{center}
\caption{The change of the Comptonised component flux ($\Delta F_{\rm nthcomp}$) as a function of the change of the disk component flux (upper, $\Delta F_{\rm diskbb}$) and the change of the inner disk temperature (lower, $\Delta kT_{\rm in}$) from non-QPO to type-B QPO. The spectra of the period with and without the type-B QPO were extracted from one or two single orbits before and after the transition. The fluxes of different components and the inner disk temperature were measured from model 2. The fluxes are in units of $10^{-8}$ erg cm$^{-2}$ s$^{-1}$. The solid line represents the best-fittingting straight line ($y=-1.02x-0.04$ for the upper panel and $y=-117.89x+0.42$ for the lower panel) to the data.}
\label{fig:flux_correlation}
\end{figure}

From epoch NQ to BQ, the inner disk temperature decreased marginally in all cases for both models 1 and 2. 
The behaviour of the {\sc diskbb} normalisation, $N_{\rm diskbb}$, which is related to inner disk radius ($N_{\rm diskbb} \propto R_{\rm in}^{2}$), depends on the model we used. When using model 2, $N_{\rm diskbb}$ decreased from epoch NQ to BQ in all cases, suggesting that the inner disk front is closer to the central BH during the appearance of the type-B QPO. However, when using model 1, $N_{\rm diskbb}$ increased from epoch NQ to BQ in all cases. 
{\sc simpl} outputs the sum of the unscattered seed photon spectrum and the scattered power-law component. The input seed spectrum of photons ({\sc diskbb} in our case) is multiplied by ($1-f_{\rm sc}$) before adding it back to the scattered power-law component which contains $f_{\rm sc}$ times the number of photons as in the {\sc diskbb}. Thus the fitted $N_{\rm diskbb}$ is not the one that is directly measured, and the model effectively assumes that the disk is screened by the Comptonised region, with only ($1-f_{\rm sc}$) leaking through.
The increase in $N_{\rm diskbb}$ from epoch NQ to BQ might be artificially induced to account for the drop in direct photons due to the rise in $f_{\rm sc}$.
In an attempt to make the disk normalisation independent of this underlying assumption made by {\sc simpl}, we used an alternative set up by fitting the data with the model {\sc tbnew*(simpl*diskbb+constant*diskbb+gaussian)}. The disk normalisation and temperature were linked and the scattered fraction of the {\sc simpl} model was fixed at unity. In this way, the disk normalisation and the {\sc simpl} component are not tied, and the {\sc constant} will allow the {\sc diskbb} to vary freely.
The best-fitting spectral parameters of our modelling are reported in Table \ref{tab:phil}. We found similar results as those reported for model 1, except that $N_{\rm diskbb}$ for both epoch BQ and NQ is much lower than what we found using model 1.

Back to our results of using models 1 and 2, we found that the photon index does not change significantly between epoch BQ and NQ. 
The best-fitting $\Gamma$ obtained is consistent between the two epochs within errors, except in the case of Obs \#1, where the $\Gamma$ is slightly larger for epoch BQ. 
However, the scattered fraction changed significantly by 26\%, 30\%, 36\%, 15\% for the four cases, respectively.
This result suggests that the increased hardening is related to an increase in normalisation of the hard component (e.g., due to an increase in scattered fraction of the Comptonised component), rather than a change in slope.
% This result appears to be at odds with the fact that a stronger hard component can be seen during the appearance of the type-B QPO, resulting in an overall spectral hardening. 
%
In addition, we note that the $\Gamma$ we measured is higher ($\Gamma \approx 3-3.5$) than the expected value ($\Gamma \approx 2.2-3$) typically observed during the SIMS in other BHBs \citep[e.g.,][]{Remillard2006}. This is possibly because the photon index cannot be well constrained due to the narrow bandpass of {\it NICER}.

The Fe K$\alpha$ emission line was weak during the period we analysed here. The contribution of the line to the total flux is $\lesssim0.2\%$ in all cases. The parameters of the {\sc gaussian} component do not show substantial variations from epoch NQ to BQ. Linking the {\sc gaussian} component between the two epochs does not worsen the fits. Fitting the iron line with more complicated models is beyond the scope of this paper, and does not affect the main conclusions of this work.

A key difference we found when comparing the BQ and NQ spectra is in the fluxes of the disk and Comptonised components.
From epoch NQ to BQ, the disk component flux decreased, while the Comptonised component flux increased in all cases. This can be inferred from the rise in scattered fraction ($f_{\rm sc}$) when using model 1, meaning that a larger fraction of seed photons are scattered into the Comptonised component during the appearance of the type-B QPO.
In Fig. \ref{fig:flux_correlation}, we plot the change of the Comptonised component flux ($\Delta F_{\rm nthcomp}$), as a function of the change of the disk component flux  ($\Delta F_{\rm diskbb}$) and the change of the inner disk temperature ($\Delta kT_{\rm in}$) from epoch NQ to BQ.
We found that $\Delta F_{\rm nthcomp}$ is anti-correlated with $\Delta F_{\rm diskbb}$ and $\Delta kT_{\rm in}$.

\begin{figure}
\begin{center}
\resizebox{\columnwidth}{!}{\rotatebox{0}{\includegraphics[clip]{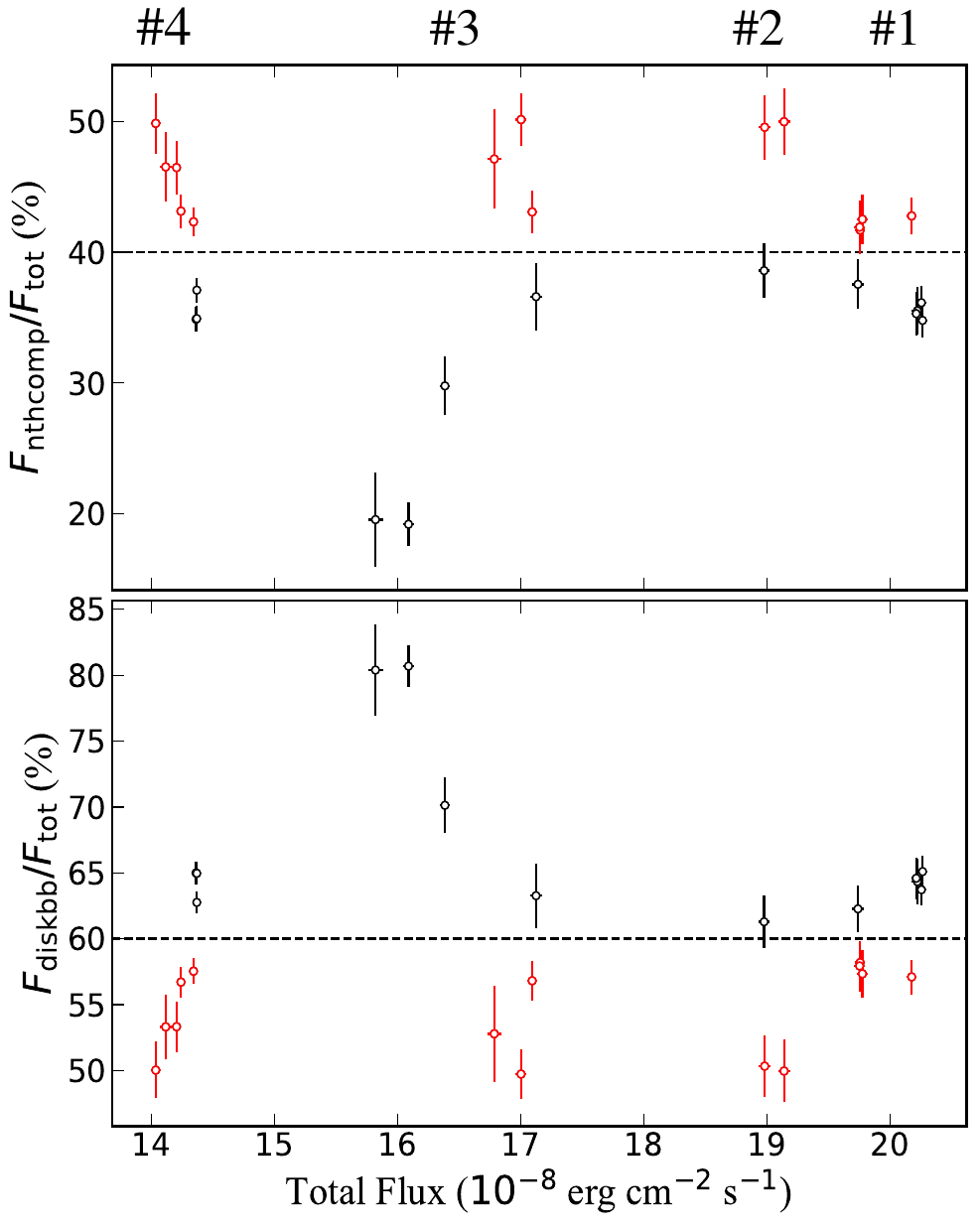}}}
\end{center}
\caption{Flux fraction of the Comptonised component ($F_{\rm nthcomp}/F_{\rm total}$, upper) and the disk component ($F_{\rm diskbb}/F_{\rm total}$, lower) as a function of the total flux for each orbit of the four observations in which a fast appearance/disappearance of a type-B QPO was observed. Each data point represents the result of one or two single orbits. The red (black) points represent the orbits with (without) a type-B QPO. Note that some of the data points are overlapped.}
\label{fig:fraction_total}
\end{figure}

\begin{figure*}
    \begin{center}
        \resizebox{2.2\columnwidth}{!}{\rotatebox{0}{\includegraphics[clip]{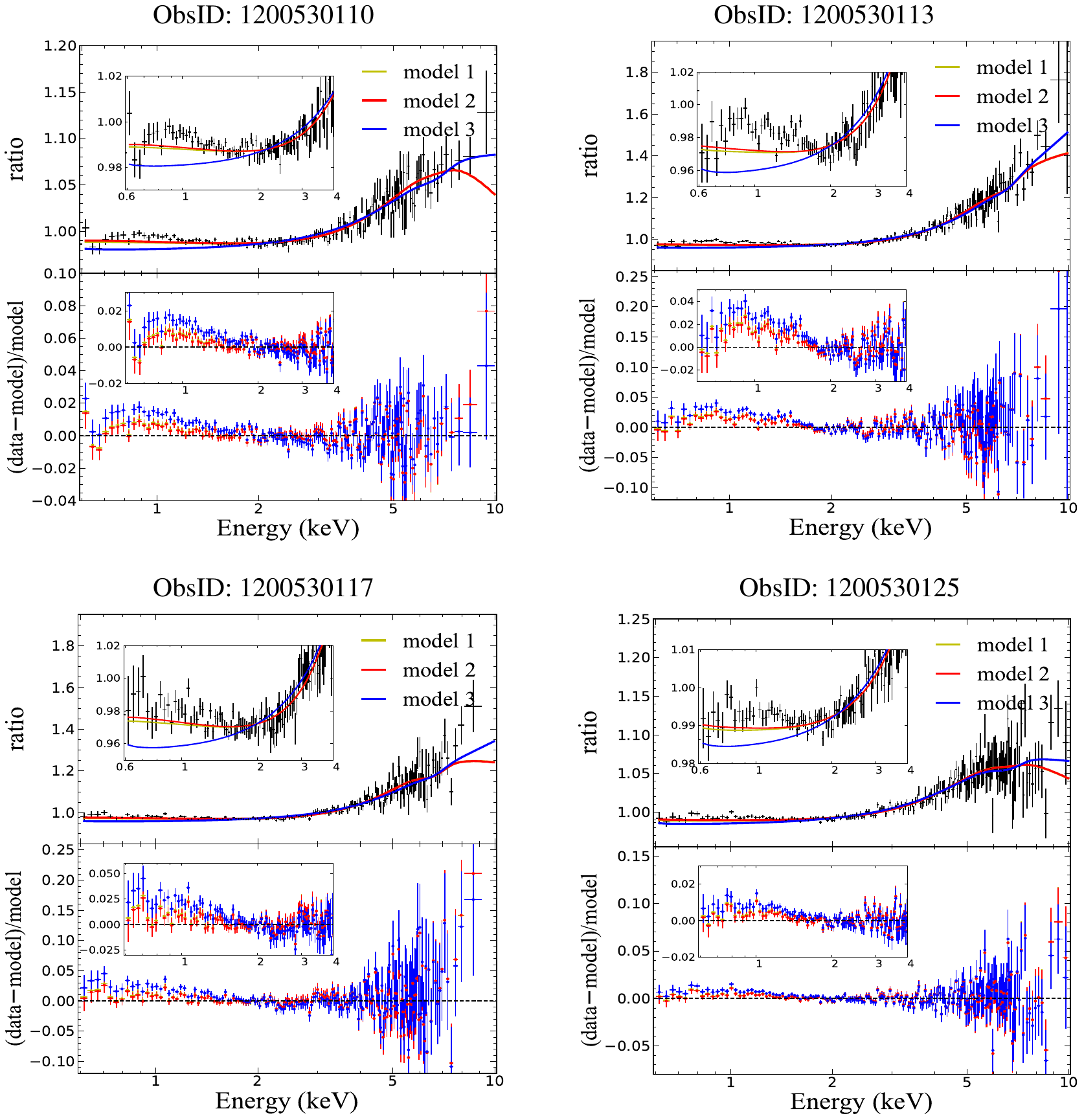}}}
    \end{center}
    \caption{\textit{Upper panels:} data-points represent the ratio between the spectra with and without the type-B QPO. 
%    
  %  The best-fitting models of the spectrum with the type-B QPO and that without the QPO. 
  In a similar way, yellow, red and blue lines correspond to the ratio of the best-fitting models for the spectra with and without the QPO for our models 1, 2 and 3, respectively. Insets show a zoom to the low-energy region. 
  %
  %The spectral ratio shown in the low panel of Fig. \ref{fig:spectra_orbit} is plotted. Lower panel:
  \textit{Lower panels:}  the ratio {\sc (DATA$-$MODEL)/MODEL} for models 1, 2 and 3 (yellow, red and blue, respectively), where {\sc DATA} and {\sc MODEL} are, respectively, the ratios of the spectra and models plotted in the upper panels. The error bars in the lower panels are underestimated, as we do not consider the errors in the model (i.e. we assume the models are perfect). Insets show a zoom to the low-energy region. Note that in this figure models 1 and 2 are overlapped.}
    \label{fig:model}
\end{figure*}

To find out which set of spectral parameters are responsible for the spectral difference, we fitted the BQ and NQ spectra of each observation simultaneously with all the parameters tied at first. Then, we allowed the parameters of the Comptonised, disk, and {\sc gaussian} components free one by one, and studied the change in $\chi^{2}/{\rm dof}$. The resulting $\chi^{2}/{\rm dof}$ is shown in Table \ref{tab:joint_fits_orbit}.
As expected, if all the parameters were tied, then we obtained very poor spectral fits with $\chi^{2}/{\rm dof}\gtrsim1.9$ (sometimes $\chi^{2}/{\rm dof}>3$).
Allowing the parameters of the Comptonised component to vary between epochs BQ and NQ brings significant improvement to the fits. In general, this produces an improvement of $\Delta\chi^{2}\gtrsim300$ through the addition of two free parameters. In all cases, freeing $\Gamma$ first results in a better fit than freeing $f_{\rm sc}$ (model 1) or $N_{\rm nthcomp}$ (model 2) first.
However, only untying the parameters of the Comptonised component is not enough to model the overall spectral difference. Variations in the {\sc diskbb} component are also required. In model 1, allowing $kT_{\rm in}$ and $N_{\rm diskbb}$ to be free improves the fits by $\Delta\chi^{2}\gtrsim90$; while in model 2, this improves the fits by $\Delta\chi^{2}\gtrsim30$. The two parameters of {\sc diskbb} were highly degenerate. It is hard to distinguish the parameter that leads to the changes. When using model 1, freeing $kT_{\rm in}$ first results in a slightly lower reduced $\chi^{2}$ than freeing $N_{\rm diskbb}$ first; in contrast, when using model 2, freeing $N_{\rm diskbb}$ first is better than freeing $kT_{\rm in}$ first.
Allowing the parameters of the {\sc gaussian} component to be free did not improve the fits.

From our spectral analysis above, we found that the appearance of the type-B QPO is related to a change in the relative contribution of the disk and Comptonised emission.
In Fig. \ref{fig:fraction_total}, we show the fraction of the Comptonised component flux to the total flux ($F_{\rm nthcomp}/F_{\rm total}$, upper) and the fraction of the disk component flux to the total flux ($F_{\rm diskbb}/F_{\rm total}$, lower) against the total flux. In this plot, each data point represents the result from one or two single orbits. The red/black points mark the orbits with/without the type-B QPO. 
We found a very clear threshold in $F_{\rm nthcomp}/F_{\rm total}$ or $F_{\rm diskbb}/F_{\rm total}$ which is related to the appearance of the type-B QPO. When $F_{\rm nthcomp}/F_{\rm total}\gtrsim40\%$ (or $F_{\rm diskbb}/F_{\rm total}\lesssim60\%$), the type-B QPO was observed. However, when $F_{\rm nthcomp}/F_{\rm total}\lesssim40\%$ (or $F_{\rm diskbb}/F_{\rm total}\gtrsim60\%$), the type-B QPO disappeared.

\subsubsection{Fit with a two-component Comptonisation model}

\cite{Garcia2021} found that the energy-dependent fractional rms and phase lags of the type-B QPO in \target\ can be well modelled with a two-component Comptonisation model. Given that Figs. \ref{fig:spectra_average} and \ref{fig:spectra_orbit} show that the emission above 2 keV is stronger when the type-B QPO appears, then it is natural to consider that either the number of spectral components does not change (but only their parameters), or that the appearance of the type-B QPO is related to the appearance of an additional spectral component.
%
%If we assume that type-B QPO spectrum contains the same hard component that is present in the non-QPO spectrum, then the QPO is expected to be associated with an additional hard component which only contribute to the type-B QPO spectra.
%
The first scenario was studied in the previous section. In this section, we further investigate the possibility that the type-B QPO originates from an additional Comptonised component.

For each observation, we fitted the BQ and NQ spectra simultaneously with a two-component Comptonisation model:

\vspace{0.25cm}

\textbf{Model 3: }  
{\sc tbnew*(constant*(diskbb+nthcomp$_{1}$+}\\ 
{\sc gaussian)+nthcomp$_{2}$)}.

\vspace{0.25cm}

In this model, we assume that the {\sc diskbb}, {\sc nthcomp$_1$} and {\sc gaussian} components are present in the spectra of both epochs, and these components have the same spectral shape with all parameters linked. The only difference is in their fluxes which is accounted for by using a {\sc constant}. The {\sc constant} was fixed at unity for epoch NQ and was allowed to vary for epoch BQ.
The {\sc nthcomp$_2$} component is assumed to be associated with the QPO and only contributes to the spectrum of epoch BQ. The normalisation of this component, $N_{\rm nthcomp2}$, for epoch NQ was fixed at zero. 
In both {\sc nthcomp$_1$} and {\sc nthcomp$_2$}, the seed photon temperature, $kT_{\rm bb}$, was linked to the inner disk temperature of the {\sc diskbb} component, $kT_{\rm in}$. The electron temperature was fixed at 1000 keV. 
The best-fitting spectral parameters for model 3 are presented in Table \ref{tab:two_nthcomp}.
Model 3 provides an slightly worse fit than model 2 for Obs. \#1, \#2 and \#3, and an approximately equivalent fit for Obs. \#4.

In model 3, compared with the same setting without the {\sc nthcomp$_2$} component, adding this component for epoch BQ reduces the $\chi^{2}$ significantly by $\Delta\chi^{2}>400$ for 2 degrees of freedom (dof) less, which yields an F-test probability of $\ll 10^{-10}$. At the same time, the value of $N_{\rm nthcomp2}$ measured is more than 4.6$\sigma$ different than zero\footnote{the best-fitting $N_{\rm nthcomp2}$ divided by its 1-$\sigma$ error.}. From all these we conclude that this component is significantly required for the BQ spectrum.
The value of $N_{\rm nthcomp2}$ for epoch NQ was fixed at zero at first. If we then let it be free, the best-fitting value of this parameter is consistent with zero, which indicates that this component is not needed in the NQ spectrum.

The disk temperature and normalisation measured from model 3 is similar to that of model 2 for epoch NQ.
In Obs \#2 and \#3, where the spectral ratio above 2 keV is much larger than that of the other two cases, the photon index of {\sc nthcomp$_2$} is significantly smaller than that of {\sc nthcomp$_1$}, suggesting that the Comptonised component associated with the QPO is harder.  While in Obs. \#1 and \#4, where the spectral ratio above 2 keV is smaller, the photon index is consistent between the two Comptonised components within errors.

In Fig. \ref{fig:model} we show the ratio between the best-fitting models of the BQ and NQ spectra for each of the models we used.
It can be seen that the two-component Comptonisation model (model 3) fits the spectral ratio better at high energy bands.
However, none of the models we used can fully account for the small difference of the spectra below 2 keV, even if we allowed the disk temperature and normalisation to vary between epoch BQ and NQ.
The small difference appears systematically in all four observations, and it is  likely  real.
We note that \nicer\ spectra shows calibration-related instrumental residuals below 2 keV \citep[see, e.g.,][]{Ludlam2018}. However, the spectral ratios we have shown in the lower panel of Fig. \ref{fig:spectra_average} and \ref{fig:spectra_orbit} are not affected by the instrumental residuals.

\section{Discussion}

Fast transitions in X-ray variability are rare events. In the past two decades there have only been reports for three sources, with a relatively large number of detections in XTE J1859+226 \citep{Casella2004, Sriram2013}, H1743$-$322 \citep{Sriram2021} and GX 339$-$4 \citep{Nespoli2003,Motta2011}.
While the small number of detections in other BHBs is probably intrinsic to the systems, it is also partly due to the limitation on the observation coverage. The multiple detection of the fast transition between non-QPO and type-B QPO in \target\ has allowed us to perform the first systematic spectral-timing analysis of such a transition in a source, and the first study with spectral information in the 0.5--2 keV range. 
Our main results can be summarized as:

\begin{itemize}
    \item the overall spectral hardness is slightly higher (harder) when the type-B QPO was detected;
    \item the main spectral difference between the period with and without the type-B QPO was seen at energies above 2 keV, and it is very small below 2 keV;
    \item during the transition from non-QPO to type-B QPO, both the disk and Comptonised fluxes changed significantly and they varied in an anti-correlated way. However, the total flux of the source in the 0.5--12 keV band remained relatively stable during the transition. 
\end{itemize}

\subsection{Spectral difference above and below 2 keV}

\citet{Stevens2016} studied the type-B QPO in GX 339$-$4 using phase-resolved spectroscopy. They found that the disk and power-law emission varied with $\sim$1.4 per cent and $\sim$25 per cent rms, respectively, in the QPO cycle. The variability of the disk flux is very small, similar to the value of the QPO fractional rms we measured at energies below 2 keV (about 0.9 per cent, see Fig. \ref{fig:energy_dependence}).
Based on a systematic study of the type-B QPO in several BHBs, \citet{Gao2017} confirmed a positive correlation between the type-B QPO frequency and the Comptonised component flux, which was first reported in GX 339$-$4 by \citet{Motta2011}.

By comparing the spectra of the period with and without the type-B QPO, we found that the main spectral difference comes from energies above 2 keV where the emission is dominated by the Comptonised component. 
Our spectral fits confirmed that the variations of the Comptonised component parameters are responsible for the main spectral difference.
These results are consistent with the findings of  \citet{Stevens2016} and \citet{Gao2017} that the Comptonised component dominates the type-B QPO emission.

At low energies (below 2 keV), the spectral difference is small. The spectral ratio below 2 keV is not constant within errors, indicating a change in the flux as well as the spectral shape.
From our joint fits with models 1 and 2, we also found that both the disk and Comptonised components are required to vary during the transition, although the variations of the {\sc diskbb} parameters are very small compared to the changes of the Comptonised component.  
In the lower panels of Fig.~\ref{fig:model} we show that while the models we used here fit the data well, they still fail to correctly fit an excess at energies below 2 keV. This excess appears systematically in all four observations, and they are only $\sim$2--4 per cent deviations from the models we used. 
We tried to model that excess with another {\sc blackbody} component, however the model did not converge and the fits became unstable. The reason why we are unable to constrain this small excess below 2 keV is probably related to the lack of broad-band coverage (which prevent us to fit the full reflection models), and the fact that $\sim$2 per cent excess is at the limit of the current NICER calibration, and therefore it is unclear how well this excess can be constrained.
However, we note that the small excess should not affect the best-fitting spectral parameters and the threshold in the relative contribution of the disk and Comptonised emission significantly.

\subsection{Flux variation during the transition and the transient nature of the type-B QPO}

\citet{Nespoli2003} reported a fast appearance of a 6-Hz type-B QPO in GX 339$-$4 with an increase in source flux. The PDS of the period without the type-B QPO showed a broad peak at around 7 Hz, which is likely to be a type-A QPO. During the transition from the potential type-A to type-B QPO, the disk flux decreased by 9 per cent, while the power-law flux increased by 30 per cent.
Fast transitions between type-A and type-B QPO were also detected in XTE J1859+226 \citep{Casella2004,Sriram2013} and XTE J1817$-$330 \citep{Sriram2012} with {\it RXTE}.
We note that such transitions are often associated with a sudden change in source flux. In the two cases of XTE J1859+226, both the disk and power-law fluxes decreased in one case (ObsID: 40124-01-13-00) and increased in the other case (ObsID: 40124-01-14-00) from type-A to type-B QPO. In the case of XTE J1817$-$330, the period with the type-B QPO has a higher flux than that with the type-A QPO, and the flux change is due to the power-law component (the disk flux does not change).
Type-C to/from type-B QPO transitions have been observed in a few sources \citep[e.g.,][]{Casella2004,Motta2011,Homan2020,Sriram2021}, which correspond to the state transition between the HIMS and SIMS. These transitions also involve a flux change, and the type-B QPOs are always present at a higher flux level.
\citet{Xu2019} detected a sudden turn-on of a transient type-C QPO in Swift J1658.2$-$4242 with a rapid flux drop. During the transition, both the disk and power-law fluxes decrease. 
Similar transient type-C QPOs were also found in three {\it RXTE} observations of H1743$-$322 \citep{Sriram2021}. However, no consistent flux change was seen during these transitions. The turn-on of the type-C QPO was accompanied by a flux rise in ObsID 80146-01-07-00, and a flux drop in ObsID 80146-01-13-00 and 80146-01-50-00.

\citet{Sriram2013} reported a non-QPO to type-B QPO transition observed in XTE J1859+226 (ObsID: 40122-01-01-00). 
All the parameters except the photon index are required to vary when fitting the spectra of the period with and without the QPO with a disk and a power-law component.
More recently, \citet{Sriram2021} found a transient 4.5-Hz type-B QPO in three {\it RXTE} observations of H1743$-$322.
In all those cases, the disk flux decreased while the power-law flux increased during the transition from non-QPO to type-B QPO, similar to what we found in \target.

As shown in Fig. \ref{fig:lc}, the 0.5--12 keV count rate (total flux) of \target\ remained more or less constant from non-QPO to type-B QPO. 
In addition, we found that both the disk and Comptonised component fluxes change significantly during the transition, and the change of the disk component flux (or $kT_{\rm in}$) is anti-correlated with that of the Comptonised component flux. 
It is worth mentioning that the flux change can be inferred from the spectral ratio shown in Fig. \ref{fig:spectra_average} and \ref{fig:spectra_orbit}, which is model independent.
Our results reveal the transient nature of the type-B QPO: the type-B QPO is associated with a redistribution of accretion power between the disk and Comptonised emission. Compared with the period without the QPO, a larger fraction of accretion power is dissipated into the Comptonised region than in the disk when the type-B QPO appears.

From Fig. \ref{fig:lc}, we found that the orbits with the type-B QPO always correspond to a slightly higher hardness ratio within each observation. However, the hardness value corresponding to the transition is different between observations. From Fig. \ref{fig:fraction_total}, we can see that it is the relative contribution of the disk and Comptonised emission that determines the appearance of the QPO, rather than the value of the hardness ratio. A very sharp threshold in the flux fraction of the disk/Comptonised component was observed. The type-B QPO was only detected when the contribution of the Comptonised emission to the total flux is above $\sim$40 per cent in the 0.6--10 keV band. Below this threshold, the type-B QPO switched off.
At present, we do not know if the critical value is the same in all BHBs, or whether source properties determine this critical value.

\subsection{Origin of the type-B QPO}

Although no comprehensive model has been proposed for the type-B QPOs, there is growing evidence suggesting that they may originate in a jet-like structure.  
The most striking observational evidence is the fact that the launch of discrete jet ejection (or the resulting radio flare) occurs close in time to the appearance of the type-B QPO, which has been found in a few BHBs \citep{Soleri2008,Fender2009,Russell2019A,Homan2020,Russell2020}.
However, the precise association between the ejection event and the type-B QPO is not clear, as the type-B QPO sometimes appears before and sometimes after the inferred jet ejection. The main obstacle of these studies is the gaps in the X-ray/radio coverage. 
The strongest evidence to date was provided by  \citet{Homan2020} who found that the start time of the radio flare is $\sim$2--2.5 hr after the appearance of the type-B QPO in MAXI J1820+070.
In addition, \citet{Homan2020} found that the appearance of the type-B QPO coincides with the start of a brief X-ray flare in the \nicer\ 7--12 keV band. The X-ray flare can possibly be due to the contribution to the X-ray flux from the jet.
In \target, a strong radio flare was observed on MJD 58523 \citep{Carotenuto2021}, which is close to the time when the type-B QPOs started to be detected (MJD 58522.6, \citetalias{Zhang2020a}). The radio flare is likely to be associated with a discrete jet ejection with its best inferred ejection date $t_{\rm ej} = {\rm MJD}~58518.9 \pm 2.4$ \citep{Carotenuto2021}. This suggests a strong connection between the radio flare (or jet ejection) and the type-B QPOs we observed.

The jet-based origin of the type-B QPO was also supported by the fact that the type-B QPOs are stronger in low-inclination (face-on) sources \citep{Motta2015}.
To explain the small amplitude of the disk varations and its phase relation to the power-law emission, \citet{Stevens2016} proposed that the Comptonised region producing the type-B QPO has a relatively large scale-height, such as the base of the jet. 
\citet{Stevens2016} also found that the power-law photon index varies between $\sim$2.3 and $\sim$2.6 within the QPO cycle, which can be quantitatively explained by a precessing jet model \citep{Kylafis2020}. Such a precessing jet has already been seen in 3D general relativistic magnetohydrodynamic simulations \citep{Liska2018}.

As shown in Fig. \ref{fig:spectra_average} and \ref{fig:spectra_orbit}, a stronger Comptonised component was seen when the type-B QPO was detected. Although we cannot rule out the one Comptonised component models, our successful spectral fit with a two-component Comptonised model suggests that the increased Comptonised emission may come from an additional component that is only present when the type-B QPO appears. Given its direct association with the type-B QPO, this component may be related to the jet.

%%%%% ADDED BY FG %%%%%
The energy-dependent fractional rms and phase lags of the type-B QPO in \target\ have been explained by a two-component Comptonisation model \citep{Garcia2021}. In this model, each Comptonisation component dominates the variability observed in the QPO in different energy ranges. Considering the dilution caused by the disk-blackbody soft-thermal emission, the model indicates that a {\em large} component ($\sim$550~$R_g$) dominates the low-amplitude variability ($\lesssim5\%$) below 4--5~keV, whereas above 4--5~keV, where the amplitude variability is larger ($\sim$5--15\%), a {\em small} component ($\sim$20~$R_g$) prevails. Significant illumination of the accretion disk by an extended Comptonised region is also expected in jet models \citep{Reig2021}.

If such a large Comptonised region is present, a narrow Fe K line would be expected in the spectrum if this component illuminates the outer parts of the disk, together with a broad line produced by the small component illuminating the inner parts of the disk. Such an effect was found by \citet{GarciaJ2019} during the 2017 failed-transition outburst of GX~339--4. In that work, the authors showed that the spectral fits are significantly improved by invoking a dual-lampost model, where the lampost component that is farthest away has a height of 500--700~$R_g$. In our spectra we clearly see a broad Fe K line. To check the possibility of the presence of a narrow line, we combined all the BQ segments (with the type-B QPO active) and added a narrow Gaussian with width fixed at 0.01 and 0.1~keV to model 1. We obtained 3-$\sigma$ upper limits of 2.5 and 4.7~eV, respectively, to the presence of such a narrow Fe line. We then performed simulations using a combination of \texttt{relxill} and \texttt{xillver} models \citep{xillver}, corresponding to a corona at 20 and 500~$R_g$ above the disk, respectively, and found consistent upper limits to such a narrow line. The potential narrow line connected to the appearance of the type-B QPO would strongly support such an extended Comptonisation region. On the contrary, more stringent constrains on the presence of such a narrow Fe line could strongly constrain those models.
%%%%% ADDED BY FG %%%%%

\section{Conclusion}

In Summary, in this work we have presented the first systematic spectral-timing analysis of a fast appearance/disappearance of a type-B QPO in \target\ using \nicer\ data. We measured the spectral difference between the periods with and without the QPO, and confirmed that the QPO emission is dominated by the hard component. Compared with the period without the QPO, the Comptonised emission increased and the disk emission decreased when the type-B QPO was detected. This suggests that the transient nature of the type-B QPO is related to a redistribution of accretion power between the disk and Comptonised emission. It is important to mention again that our results focus on the energy bands below 10 keV. 
Future studies using broad-band X-ray data (e.g. 0.3-100 keV) will be able to better constrain the different spectral components (specially the hard and the reflection components), and give deeper insights into the origin of the QPOs. This would be possible if similar transitions to those presented in this paper are observed simultaneously with \nicer\ and NuSTAR \citep{Harrison2013}, or \nicer\ and {\it Insight}/HXMT \citep{Zhang2020}.

\begin{landscape}
\begin{table}
\centering
\footnotesize
\caption{Best-fitting spectral parameters for models 1 and 2. The spectra of the period with and without the type-B QPO were extracted from one or two single orbits before and after the transition.}
\label{tab:best_fits_orbit}
\begin{tabular}{lcccccccccccccccc}
\hline\hline
Component  & Parameters & \multicolumn{2}{c}{1200530110}   && \multicolumn{2}{c}{1200530113} 
                        &&  \multicolumn{2}{c}{1200530117} && \multicolumn{2}{c}{1200530125}\\
\cline{3-4} \cline{6-7} \cline{9-10} \cline{12-13}
           &            & type-B QPO  & non-QPO     && type-B QPO    & non-QPO  
                        && type-B QPO & non-QPO     && type-B QPO    & non-QPO \\
           &            & orbit 4     & orbit 3     && orbit 2\&3    & orbit 4 
                        && orbit 5    & orbit 4     && orbit 5       & orbit 6\&7 \\
\hline
\multicolumn{13}{c}{Model 1: \textsc{tbnew*(simpl*diskbb+gaussian)}} \\
\hline
TBNEW  &  $N_{\rm H}$ ($\times10^{22}~{\rm cm}^{-2}$) &  \multicolumn{2}{c}{$0.89 \pm 0.01$} && \multicolumn{2}{c}{$0.89 \pm 0.01$} 
                                                      && \multicolumn{2}{c}{$0.89 \pm 0.01$} && \multicolumn{2}{c}{$0.89 \pm 0.01$} \\
DISKBB &  $kT_{\rm in}$ (keV)   &  $0.693 \pm 0.004$   & $0.708 \pm 0.004$  
                                && $0.671 \pm 0.006$   & $0.694 \pm 0.005$  
                                && $0.650 \pm 0.005$   & $0.677 \pm 0.005$ 
                                && $0.640 \pm 0.004$   & $0.645 \pm 0.003$  \\
                                
       &  Norm                  &  $43441 \pm 870$    & $41048 \pm 773$   
                                && $44783 \pm 1447$   & $41307 \pm 980$ 
                                && $46000 \pm 1279$   & $41631 \pm 1197$ 
                                && $43152 \pm 1080$   & $42578 \pm 755$  \\
                                
SIMPL  &  $\Gamma$              &  $3.48 \pm 0.04$   & $3.30 \pm 0.05$  
                                && $3.37 \pm 0.06$   & $3.46 \pm 0.07$  
                                && $3.46 \pm 0.05$   & $3.37 \pm 0.10$ 
                                && $3.35 \pm 0.05$   & $3.27 \pm 0.04$  \\
                                
       &  $f_{\rm SC}$          &  $0.27 \pm 0.01$   & $0.20 \pm 0.01$ 
                                && $0.33 \pm 0.02$   & $0.23 \pm 0.02$  
                                && $0.33 \pm 0.02$   & $0.21 \pm 0.02$ 
                                && $0.26 \pm 0.01$   & $0.22 \pm 0.01$  \\
                                
GAUSSIAN &  $E_{\rm line}$ (keV)&  $6.4^a$           & $6.4^a$  
                                && $6.4^a$           & $6.4^a$                
                                && $6.4^a$           & $6.4^a$      
                                && $6.42 \pm 0.13$   & $6.48 \pm 0.08$  \\
                                
         &  $\sigma$ (keV)      &  $0.56 \pm 0.10$   & $0.62 \pm 0.12$  
                                && $0.5^a$           & $0.5^a$                 
                                && $0.5^a$           & $0.5^a$        
                                && $0.57 \pm 0.13$   & $0.49 \pm 0.09$  \\
                                
         &  Norm                &  $0.024 \pm 0.006$ & $0.030 \pm 0.007$  
                                && $0.017 \pm 0.006$ & $0.024 \pm 0.005$                  
                                && $0.016 \pm 0.004$ & $0.023 \pm 0.006$  
                                && $0.021 \pm 0.006$ & $0.018 \pm 0.004$  \\
\hline
         &  $\chi^{2}/{\rm dof}$ &  \multicolumn{2}{c}{422.2/447}  &&  \multicolumn{2}{c}{462.9/445 } 
                                 && \multicolumn{2}{c}{461.9/436}  &&  \multicolumn{2}{c}{457.6/445 }\\
\hline
\multicolumn{13}{c}{Model 2: \textsc{tbnew*(diskbb+nthcomp+gaussian)}}\\
\hline
TBNEW  &  $N_{\rm H}$ ($\times10^{22}~{\rm cm}^{-2}$) &  \multicolumn{2}{c}{$0.90 \pm 0.01$} && \multicolumn{2}{c}{$0.89 \pm 0.01$}  
                                                      && \multicolumn{2}{c}{$0.90 \pm 0.01$} && \multicolumn{2}{c}{$0.89 \pm 0.01$}\\
                                                      
DISKBB &  $kT_{\rm in}$ (keV)   &  $0.707 \pm 0.003$    & $0.718 \pm 0.003$      
                                && $0.689 \pm 0.005$    & $0.706 \pm 0.004$   
                                && $0.668 \pm 0.004$    & $0.687 \pm 0.004$     
                                && $0.653 \pm 0.004$    & $0.655 \pm 0.003$   \\
                                
       &  Norm                  &  $27371 \pm 266$    & $29560 \pm 224$     
                                && $24780 \pm 541$    & $28113 \pm 372$   
                                && $25403 \pm 450$    & $29342 \pm 469$   
                                && $27940 \pm 221$    & $29764 \pm 217$   \\
                                
NTHCOMP  &  $\Gamma$            &  $3.35 \pm 0.04$    & $3.18 \pm 0.05$     
                                && $3.25 \pm 0.06$    & $3.33 \pm 0.07$   
                                && $3.34 \pm 0.05$    & $3.24 \pm 0.09$   
                                && $3.23 \pm 0.05$    & $3.16 \pm 0.04$   \\
                                
         &  Norm                &  $19.1 \pm 0.9$    & $14.5 \pm 0.8$      
                                && $21.5 \pm 1.4$    & $15.7 \pm 1.1$   
                                && $20.1 \pm 1.1$    & $13.6 \pm 1.3$    
                                && $13.8 \pm 0.8$    & $12.1 \pm 0.6$   \\
                                
GAUSSIAN &  $E_{\rm line}$ (keV)&  $6.4^a$            & $6.4^a$       
                                && $6.4^a$            & $6.4^a$            
                                && $6.4^a$            & $6.4^a$           
                                && $6.34 \pm 0.14$    & $6.45 \pm 0.09$  \\
                                
         &  $\sigma$ (keV)      &  $0.58 \pm 0.10$    & $0.64 \pm 0.12$      
                                && $0.5^a$            & $0.5^a$          
                                && $0.5^a$            & $0.5^a$         
                                && $0.66 \pm 0.14$    & $0.53 \pm 0.11$  \\
                                
         &  Norm                &  $0.026 \pm 0.006$  & $0.032 \pm 0.008$ 
                                && $0.019 \pm 0.006$  & $0.025 \pm 0.005$               
                                && $0.017 \pm 0.004$  & $0.024 \pm 0.006$  
                                && $0.026 \pm 0.007$  & $0.020 \pm 0.005$  \\
\hline
         &  $\chi^{2}/{\rm dof}$ &  \multicolumn{2}{c}{427.1/447}  && \multicolumn{2}{c}{464.7/445} 
                                 && \multicolumn{2}{c}{465.7/436}  && \multicolumn{2}{c}{463.2/445} \\ 
\hline
         &  ${F_{\rm diskbb}}^{b}$    &    11.56   &  13.34    
                                      &&   9.33    &  11.79    
                                      &&   8.36    &  10.97  
                                      &&   8.34    &  9.00   \\
                                      
         &  ${F_{\rm nthcomp}}^{b}$   &    8.59    &  6.87  
                                      &&   9.68    &  7.13   
                                      &&   8.65    &  6.10   
                                      &&   5.97    &  5.33  \\
                                      
         &  ${F_{\rm gaussian}}^{b}$  &    0.03    &  0.03   
                                      &&   0.02    &  0.03   
                                      &&   0.02    &  0.02    
                                      &&   0.03    &  0.02  \\
                                      
         &  ${F_{\rm total}}^{b}$     &    20.18   &  20.24    
                                      &&   19.03   &  18.95  
                                      &&   17.03   &  17.09    
                                      &&   14.34   &  14.35 \\
\hline
\multicolumn{13}{l}{$^a$ Fixed.} \\
\multicolumn{13}{l}{$^b$ 0.6--10 keV unabsorbed flux in units of $10^{-8}$ erg cm$^{-2}$ s$^{-1}$.}
\end{tabular}
\end{table}
\end{landscape}

\begin{table*}
\centering
\caption{Results of simultaneous spectral fits for models 1 and 2. The spectra of the period with and without the type-B QPO were extracted from one or two single orbits before and after the transition.}
\label{tab:joint_fits_orbit}
\begin{tabular}{lcccc}
\hline\hline
Parameters      &  1200530110 & 1200530113 & 1200530117 & 1200530125 \\
                &  $\chi^{2}/{\rm dof}$  &  $\chi^{2}/{\rm dof}$ & $\chi^{2}/{\rm dof}$ & $\chi^{2}/{\rm dof}$ \\
\hline
\multicolumn{5}{c}{Model 1: \textsc{tbnew*(simpl*diskbb+gaussian)}}\\
\hline
All tied                 & 847/453      & 2244/450     & 1416/441      & 1026/452 \\
Free $\Gamma$            & 599/452      & 998/449      & 989/440       & 549/451 \\
Free $f_{\rm SC}$        & 548/451      & 699/448      & 663/439       & 547/450 \\
Free $kT_{\rm in}$       & 428/450      & 469/447      & 470/438       & 459/449 \\
Free $N_{\rm diskbb}$    & 423/449      & 464/446      & 463/437       & 458/448 \\
All free                 & 422/447      & 463/445      & 462/436       & 458/445 \\
\hline
All tied                 & 847/453      & 2244/450     & 1416/441      & 1026/452 \\
Free $f_{\rm SC}$        & 680/452      & 1434/449     & 1218/440      & 609/451 \\
Free $\Gamma$            & 548/451      & 699/448      & 663/439       & 547/450 \\
Free $N_{\rm diskbb}$    & 434/450      & 475/447      & 478/438       & 463/449 \\
Free $kT_{\rm in}$       & 423/449      & 464/446      & 463/437       & 458/448 \\
All free                 & 422/447      & 463/445      & 462/436       & 458/445 \\
\hline
\multicolumn{5}{c}{Model 2: \textsc{tbnew*(diskbb+nthcomp+gaussian)}} \\
\hline   
All tied                 &  852/453     & 2247/450     &  1420/441     & 1032/452 \\ 
Free $\Gamma$            &  616/452     & 1094/449     &  1030/440     & 560/451 \\
Free $N_{\rm nthcomp}$   &  472/451     & 493/448      &  497/439      & 522/450 \\
Free $kT_{\rm in}$       &  472/450     & 493/447      &  497/438      & 520/449 \\
Free $N_{\rm diskbb}$    &  428/449     & 465/446      &  467/437      & 464/448 \\
All free                 &  427/447     & 465/445      &  466/436      & 463/445 \\
\hline
All tied                 &  852/453     & 2247/450     &  1420/441     & 1032/452 \\
Free $N_{\rm nthcomp}$   &  817/452     & 2103/449     &  1420/440     & 802/451  \\
Free $\Gamma$            &  472/451     & 493/448      &  497/439      & 522/450  \\
Free $N_{\rm diskbb}$    &  436/450     & 476/447      &  481/438      & 467/449 \\
Free $kT_{\rm in}$       &  428/449     & 465/446      &  467/437      & 464/448 \\
All free                 &  427/447     & 465/445      &  466/436      & 463/445 \\
\hline                   
\end{tabular}
\end{table*}

\begin{landscape}
\begin{table}
\centering
\footnotesize
\caption{Best-fitting spectral parameters for model 3. The spectra of the period with and without the type-B QPO were extracted from one or two single orbits before and after the transition.}
\label{tab:two_nthcomp}
\begin{tabular}{lcccccccccccccccc}
\hline\hline
Component  & Parameters & \multicolumn{2}{c}{1200530110}   && \multicolumn{2}{c}{1200530113} 
                        &&  \multicolumn{2}{c}{1200530117} && \multicolumn{2}{c}{1200530125}\\
\cline{3-4} \cline{6-7} \cline{9-10} \cline{12-13}
          &            & type-B QPO  & non-QPO     && type-B QPO    & non-QPO  
                        && type-B QPO & non-QPO     && type-B QPO    & non-QPO \\
          &            & orbit 4     & orbit 3     && orbit 2\&3    & orbit 4 
                        && orbit 5    & orbit 4     && orbit 5       & orbit 6\&7 \\
\hline
\multicolumn{13}{c}{Model 3: \textsc{tbnew*(constant*(diskbb+nthcomp$_1$+gaussian)+nthcomp$_2$)}} \\
\hline
TBNEW  &  $N_{\rm H}$ ($\times10^{22}~{\rm cm}^{-2}$) &  \multicolumn{2}{c}{$0.90 \pm 0.01$} 
                                                      && \multicolumn{2}{c}{$0.89 \pm 0.01$} 
                                                      && \multicolumn{2}{c}{$0.90 \pm 0.01$} 
                                                      && \multicolumn{2}{c}{$0.89 \pm 0.01$} \\
                                                      
CONSTANT &                      &  $0.94 \pm 0.01$             & $1^a$ 
                                && $0.90 \pm 0.01$             & $1^a$
                                && $0.89 \pm 0.01$             & $1^a$
                                && $0.94 \pm 0.01$             & $1^a$  \\

DISKBB &  $kT_{\rm in}$ (keV)   &  \multicolumn{2}{c}{$0.713 \pm 0.002$}
                                && \multicolumn{2}{c}{$0.699 \pm 0.003$}  
                                && \multicolumn{2}{c}{$0.675 \pm 0.003$} 
                                && \multicolumn{2}{c}{$0.654 \pm 0.003$}  \\
                                
      &  Norm                   &  \multicolumn{2}{c}{$29380 \pm 228$}   
                                && \multicolumn{2}{c}{$27604 \pm 414$} 
                                && \multicolumn{2}{c}{$28528 \pm 556$} 
                                && \multicolumn{2}{c}{$29733 \pm 211$}  \\
                                
NTHCOMP$_1$ &  $\Gamma$         &  \multicolumn{2}{c}{$3.27 \pm 0.04$}  
                                && \multicolumn{2}{c}{$3.44 \pm 0.06$}  
                                && \multicolumn{2}{c}{$3.45 \pm 0.07$} 
                                && \multicolumn{2}{c}{$3.19 \pm 0.03$}  \\
                                
            &  Norm             &  \multicolumn{2}{c}{$16.0 \pm 0.7$} 
                                && \multicolumn{2}{c}{$17.9 \pm 1.1$}  
                                && \multicolumn{2}{c}{$17.2 \pm 1.2$}
                                && \multicolumn{2}{c}{$12.4 \pm 0.5$}  \\
                                
GAUSSIAN &  $E_{\rm line}$ (keV)&  \multicolumn{2}{c}{$6.4^a$} 
                                && \multicolumn{2}{c}{$6.4^a$}                 
                                && \multicolumn{2}{c}{$6.4^a$}     
                                && \multicolumn{2}{c}{$6.41 \pm 0.08$}  \\
                                
         &  $\sigma$ (keV)      &  \multicolumn{2}{c}{$0.61 \pm 0.08$}  
                                && \multicolumn{2}{c}{$0.5^a$}                 
                                && \multicolumn{2}{c}{$0.5^a$}        
                                && \multicolumn{2}{c}{$0.58 \pm 0.10$}  \\
                                
         &  Norm                &  \multicolumn{2}{c}{$0.030 \pm 0.005$}  
                                && \multicolumn{2}{c}{$0.023 \pm 0.004$}                  
                                && \multicolumn{2}{c}{$0.021 \pm 0.004$}  
                                && \multicolumn{2}{c}{$0.023 \pm 0.005$}  \\

NTHCOMP$_2$ &  $\Gamma$         &  $3.27 \pm 0.19$   &    -  
                                && $2.52 \pm 0.11$   &    - 
                                && $2.76 \pm 0.14$   &    - 
                                && $3.36 \pm 0.15$   &    -  \\
                                
            &  Norm             &  $2.3 \pm 0.5$     &    -
                                && $3.5 \pm 0.4$     &    - 
                                && $3.3 \pm 0.5$     &    -  
                                && $1.9 \pm 0.3$     &    - \\     
\hline
         &  $\chi^{2}/{\rm dof}$ &  \multicolumn{2}{c}{437.0/450}  &&  \multicolumn{2}{c}{477.9/447 } 
                                 && \multicolumn{2}{c}{482.6/438}  &&  \multicolumn{2}{c}{467.5/449 }\\
\hline
         &  ${F_{\rm diskbb}}^{b}$      & 12.08  &  12.83   && 9.95   & 11.11   && 8.82  &  9.86    && 8.41   & 8.93  \\
         &  ${F_{\rm nthcomp_1}}^{b}$   & 6.97   &  7.41    && 7.02   & 7.84    && 6.51  &  7.27    && 5.09   & 5.41  \\
         &  ${F_{\rm gaussian}}^{b}$    & 0.03   &  0.03    && 0.02   & 0.02    && 0.02  &  0.02    && 0.02   & 0.02 \\
         &  ${F_{\rm nthcomp_2}}^{b}$   & 1.06   &  -       && 1.98   & -       && 1.65  &  -       && 0.81   & -     \\
         &  ${F_{\rm total}}^{b}$       & 20.14  &  20.27   && 18.97  & 18.97   && 17.00 &  17.15   && 14.33  & 14.36 \\
\hline
\multicolumn{13}{l}{$^a$ Fixed.} \\
\multicolumn{13}{l}{$^b$ 0.6--10 keV unabsorbed flux in units of $10^{-8}$ erg cm$^{-2}$ s$^{-1}$.}
\end{tabular}
\end{table}
\end{landscape}

\section*{Acknowledgements}

The authors would like to thank the anonymous referee for valuable comments. L.Z. acknowledges support from the Royal Society Newton Funds. D.A. acknowledges support from the Royal Society.  K.A. acknowledges support from a UGC-UKIERI Phase 3 Thematic Partnership (UGC-UKIERI-2017-18-006; PI: P. Gandhi). F.G. and M.M. acknowledge the research programme Athena with project number 184.034.002, which is (partly) financed by the Dutch Research Council (NWO). L.Z. also acknowledges supports from the National Program on Key Research and Development Project (Grant No. 2016YFA0400800), and the National Natural Science Foundation of China (Grant Nos. U1838111, U1838115, U1838201, U2038104, U2031205, 11733009). This work was supported by NASA through the NICER mission and the Astrophysics Explorers Program, and made use of data and software provided by the High Energy Astrophysics Science Archive Research Center (HEASARC).

\section*{Data Availability}

The data underlying this article are available in the HEASARC database.

%%%%%%%%%%%%%%%%%%%%%%%%%%%%%%%%%%%%%%%%%%%%%%%%%%



%\bibliographystyle{mnras}

% \bibliographystyle{mnras}
% \bibliography{ms}

%%%%%%%%%%%%%%%%%%%%%%%%%%%%%%%%%%%%%%%%%%%%%%%%%%

%%%%%%%%%%%%%%%%% APPENDICES %%%%%%%%%%%%%%%%%%%%%

\appendix
\section{Additional Figure and Table}

\begin{figure*}
\begin{center}
\resizebox{2\columnwidth}{!}{\rotatebox{0}{\includegraphics[clip]{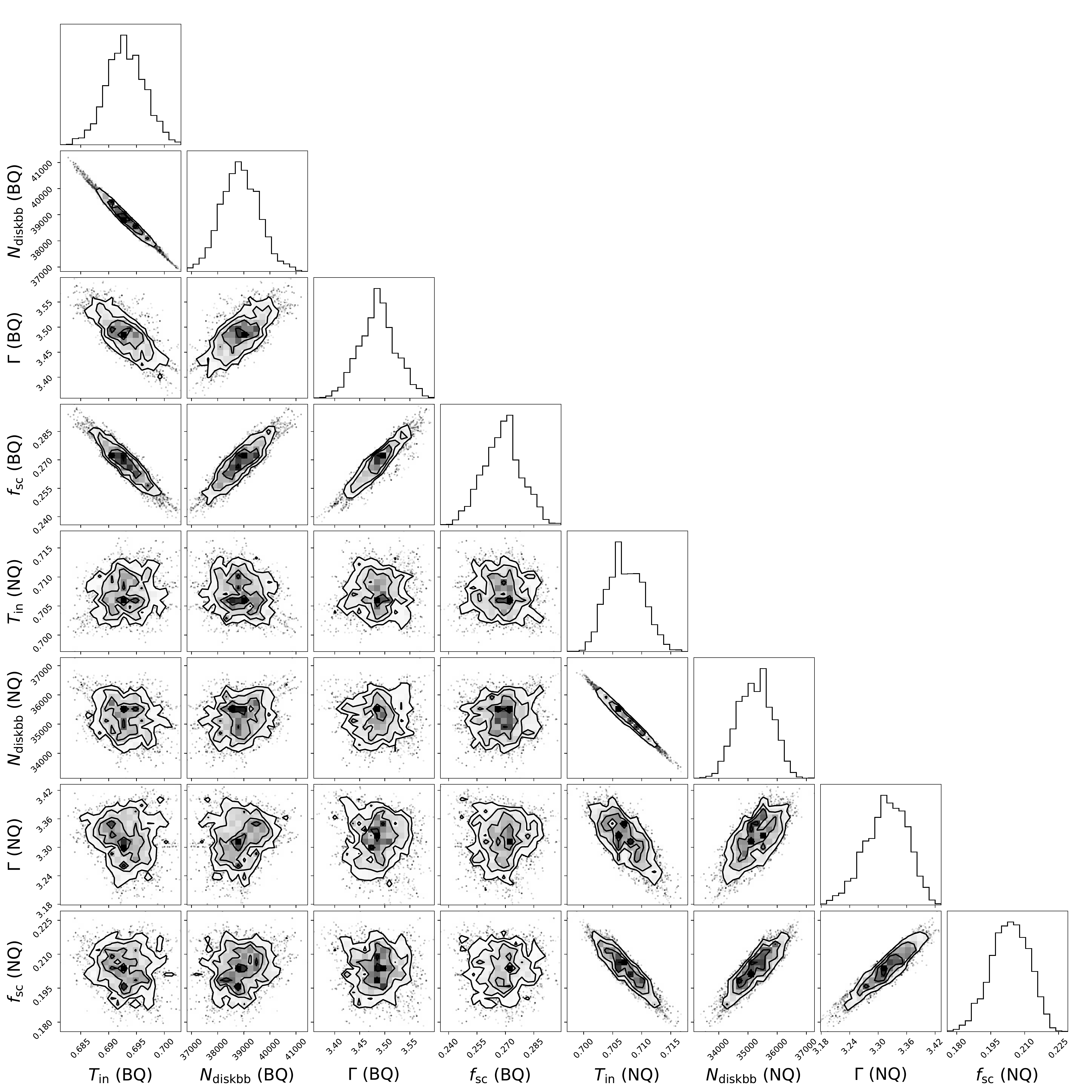}}}
\end{center}
\caption{Distributions of the main best-fitting spectral parameters fitted with model 1: \textsc{tbnew*(simpl*diskbb+gaussian)} for Obs. \#1: 1200530110.}
\label{fig:corner_simpl}
\end{figure*}

\begin{figure*}
\begin{center}
\resizebox{2\columnwidth}{!}{\rotatebox{0}{\includegraphics[clip]{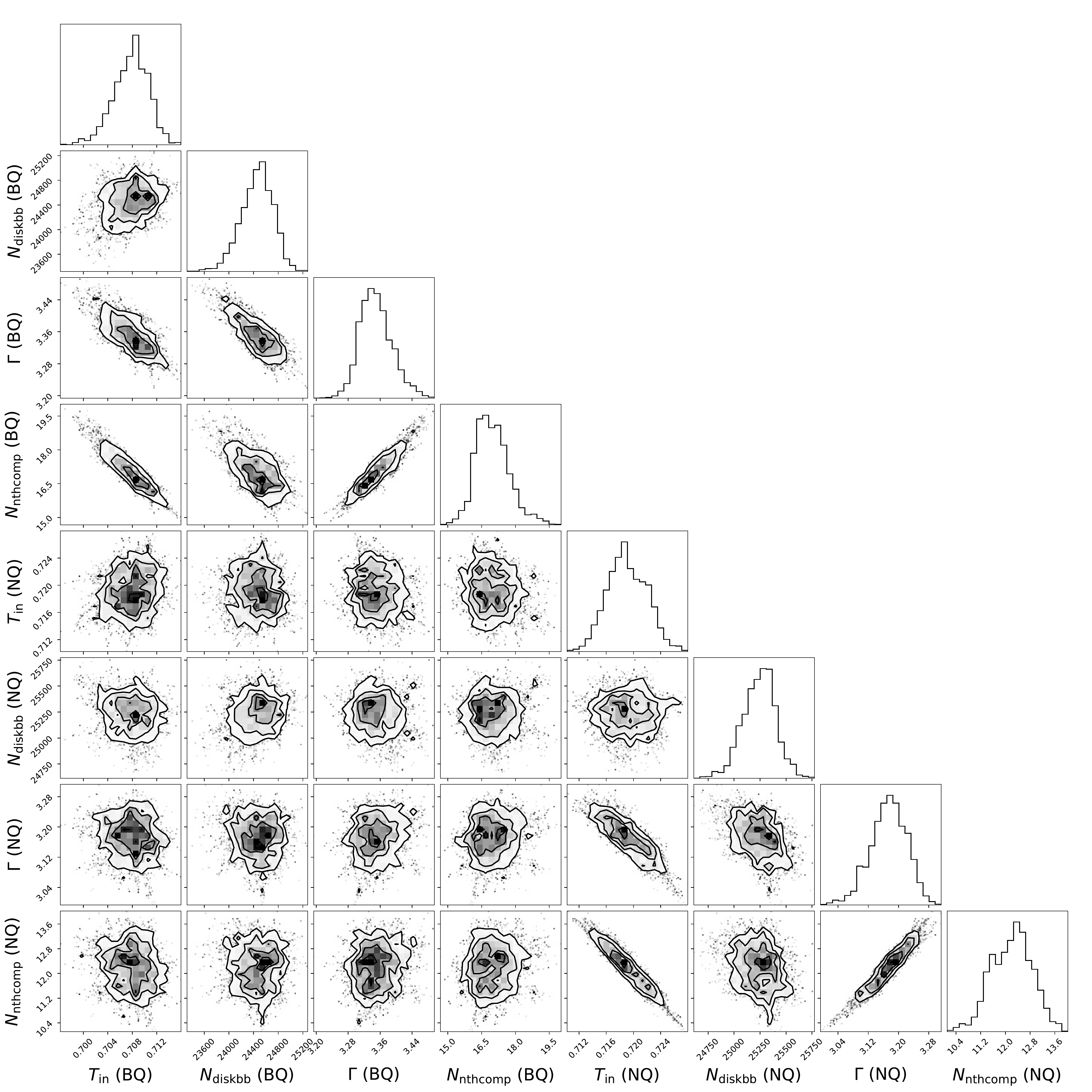}}}
\end{center}
\caption{Distributions of the main best-fitting spectral parameters fitted with model 2: \textsc{tbnew*(diskbb+nthcomp+gaussian)} for Obs. \#1: 1200530110.}
\label{fig:corner_nthcomp}
\end{figure*}

\begin{landscape}
\begin{table}
\centering
\footnotesize
\caption{Best-fitting spectral parameters for models 1 and 2. The spectra were extracted from all the orbits with and without the type-B QPO, respectively.}
\label{tab:best_fits_avg}
\begin{tabular}{lcccccccccccccccc}
\hline\hline
Component  & Parameters & \multicolumn{2}{c}{1200530110}   && \multicolumn{2}{c}{1200530113} 
                        &&  \multicolumn{2}{c}{1200530117} && \multicolumn{2}{c}{1200530125}\\
\cline{3-4} \cline{6-7} \cline{9-10} \cline{12-13}
           &            & type-B QPO  & non-QPO     && type-B QPO    & non-QPO  
                        && type-B QPO & non-QPO     && type-B QPO    & non-QPO \\
\hline
\multicolumn{13}{c}{Model 1: \textsc{tbnew*(simpl*diskbb+gaussian)}} \\
\hline
TBNEW  &  $N_{\rm H}$ ($\times10^{22}~{\rm cm}^{-2}$) &  \multicolumn{2}{c}{$0.89 \pm 0.01$} && \multicolumn{2}{c}{$0.89 \pm 0.01$} 
                                                      && \multicolumn{2}{c}{$0.89 \pm 0.01$} && \multicolumn{2}{c}{$0.89 \pm 0.01$} \\
DISKBB &  $kT_{\rm in}$ (keV)   &  $0.691 \pm 0.003$   & $0.704 \pm 0.003$  
                                && $0.671 \pm 0.005$   & $0.694 \pm 0.005$  
                                && $0.657 \pm 0.004$   & $0.678 \pm 0.003$ 
                                && $0.636 \pm 0.003$   & $0.649 \pm 0.003$  \\
                                
       &  Norm                  &  $43272 \pm 767$    & $41662 \pm 650$   
                                && $44946 \pm 1166$   & $41390 \pm 977$ 
                                && $44877 \pm 901$    & $41617 \pm 688$ 
                                && $43608 \pm 800$    & $41943 \pm 688$  \\
                                
SIMPL  &  $\Gamma$              &  $3.45 \pm 0.03$   & $3.36 \pm 0.03$  
                                && $3.33 \pm 0.05$   & $3.46 \pm 0.07$  
                                && $3.36 \pm 0.04$   & $3.44 \pm 0.07$ 
                                && $3.34 \pm 0.03$   & $3.26 \pm 0.03$  \\
                                
       &  $f_{\rm SC}$          &  $0.27 \pm 0.01$   & $0.22 \pm 0.01$ 
                                && $0.32 \pm 0.02$   & $0.23 \pm 0.02$  
                                && $0.30 \pm 0.01$   & $0.16 \pm 0.01$ 
                                && $0.27 \pm 0.01$   & $0.20 \pm 0.01$  \\
                                
GAUSSIAN &  $E_{\rm line}$ (keV)&  $6.4^a$           & $6.4^a$  
                                && $6.4^a$           & $6.4^a$                
                                && $6.4^a$           & $6.4^a$      
                                && $6.45 \pm 0.07$   & $6.40 \pm 0.06$  \\
                                
         &  $\sigma$ (keV)      &  $0.70 \pm 0.09$   & $0.63 \pm 0.06$  
                                && $0.5^a$           & $0.5^a$                 
                                && $0.5^a$           & $0.5^a$        
                                && $0.52 \pm 0.07$   & $0.61 \pm 0.07$  \\
                                
         &  Norm                &  $0.028 \pm 0.005$ & $0.031 \pm 0.004$  
                                && $0.012 \pm 0.005$ & $0.024 \pm 0.005$                  
                                && $0.012 \pm 0.003$ & $0.016 \pm 0.003$  
                                && $0.018 \pm 0.003$ & $0.021 \pm 0.003$  \\
\hline
         &  $\chi^{2}/{\rm dof}$ &  \multicolumn{2}{c}{440.8/447}  &&  \multicolumn{2}{c}{456.6/447 } 
                                 && \multicolumn{2}{c}{427.4/449}  &&  \multicolumn{2}{c}{452.0/445 }\\
\hline
\multicolumn{13}{c}{Model 2: \textsc{tbnew*(diskbb+nthcomp+gaussian)}}\\
\hline
TBNEW  &  $N_{\rm H}$ ($\times10^{22}~{\rm cm}^{-2}$) &  \multicolumn{2}{c}{$0.90 \pm 0.01$} && \multicolumn{2}{c}{$0.89 \pm 0.01$}  
                                                      && \multicolumn{2}{c}{$0.90 \pm 0.01$} && \multicolumn{2}{c}{$0.89 \pm 0.01$}\\
                                                      
DISKBB &  $kT_{\rm in}$ (keV)   &  $0.705 \pm 0.003$    & $0.715 \pm 0.003$      
                                && $0.688 \pm 0.004$    & $0.706 \pm 0.004$   
                                && $0.671 \pm 0.003$    & $0.684 \pm 0.003$     
                                && $0.649 \pm 0.003$    & $0.658 \pm 0.003$   \\
                                
       &  Norm                  &  $27324 \pm 200$    & $29156 \pm 165$     
                                && $25458 \pm 385$    & $28121 \pm 374$   
                                && $26662 \pm 268$    & $32433 \pm 239$   
                                && $27459 \pm 189$    & $30104 \pm 185$   \\
                                
NTHCOMP  &  $\Gamma$            &  $3.33 \pm 0.03$    & $3.24 \pm 0.03$     
                                && $3.22 \pm 0.05$    & $3.33 \pm 0.07$   
                                && $3.25 \pm 0.03$    & $3.32 \pm 0.06$   
                                && $3.23 \pm 0.03$    & $3.16 \pm 0.03$   \\
                                
         &  Norm                &  $18.9 \pm 0.7$    & $15.6 \pm 0.6$      
                                && $21.0 \pm 1.1$    & $15.8 \pm 1.1$   
                                && $18.4 \pm 0.7$    & $10.2 \pm 0.6$    
                                && $14.7 \pm 0.5$    & $11.4 \pm 0.5$   \\
                                
GAUSSIAN &  $E_{\rm line}$ (keV)&  $6.4^a$            & $6.4^a$       
                                && $6.4^a$            & $6.4^a$            
                                && $6.4^a$            & $6.4^a$           
                                && $6.42 \pm 0.07$    & $6.37 \pm 0.06$  \\
                                
         &  $\sigma$ (keV)      &  $0.73 \pm 0.09$    & $0.65 \pm 0.06$      
                                && $0.5^a$            & $0.5^a$          
                                && $0.5^a$            & $0.5^a$         
                                && $0.56 \pm 0.08$    & $0.63 \pm 0.07$  \\
                                
         &  Norm                &  $0.031 \pm 0.005$  & $0.032 \pm 0.004$ 
                                && $0.013 \pm 0.005$  & $0.025 \pm 0.005$               
                                && $0.013 \pm 0.003$  & $0.016 \pm 0.003$  
                                && $0.020 \pm 0.004$  & $0.022 \pm 0.004$  \\
\hline
         &  $\chi^{2}/{\rm dof}$ &  \multicolumn{2}{c}{447.0/447}  && \multicolumn{2}{c}{458.9/447} 
                                 && \multicolumn{2}{c}{432.1/449}  && \multicolumn{2}{c}{457.8/445} \\ 
\hline
         &  ${F_{\rm diskbb}}^{b}$    &    11.41   &  12.88    
                                      &&   9.55    &  11.77    
                                      &&   8.99    &  11.88  
                                      &&   7.96    &  9.26   \\
                                      
         &  ${F_{\rm nthcomp}}^{b}$   &    8.53    &  7.27  
                                      &&   9.53    &  7.15   
                                      &&   8.09    &  4.50   
                                      &&   6.31    &  5.06  \\
                                      
         &  ${F_{\rm gaussian}}^{b}$  &    0.03    &  0.03   
                                      &&   0.01    &  0.03   
                                      &&   0.01    &  0.02    
                                      &&   0.02    &  0.02  \\
                                      
         &  ${F_{\rm total}}^{b}$     &    19.97   &  20.18    
                                      &&   19.09   &  18.95  
                                      &&   17.09   &  16.40    
                                      &&   14.29   &  14.34 \\
\hline
\multicolumn{13}{l}{$^a$ Fixed.} \\
\multicolumn{13}{l}{$^b$ 0.6--10 keV unabsorbed flux in units of $10^{-8}$ erg cm$^{-2}$ s$^{-1}$.}
\end{tabular}
\end{table}
\end{landscape}

\begin{landscape}
\begin{table}
\centering
\footnotesize
\caption{Best-fitting spectral parameters for model: {\sc tbnew*(simpl*diskbb+constant*diskbb+gaussian)}. The spectra of the period with and without the type-B QPO were extracted from one or two single orbits before and after the transition.}
\label{tab:phil}
\begin{tabular}{lcccccccccccccccc}
\hline\hline
Component  & Parameters & \multicolumn{2}{c}{1200530110}   && \multicolumn{2}{c}{1200530113} 
                        &&  \multicolumn{2}{c}{1200530117} && \multicolumn{2}{c}{1200530125}\\
\cline{3-4} \cline{6-7} \cline{9-10} \cline{12-13}
           &            & type-B QPO  & non-QPO     && type-B QPO    & non-QPO  
                        && type-B QPO & non-QPO     && type-B QPO    & non-QPO \\
           &            & orbit 4     & orbit 3     && orbit 2\&3    & orbit 4 
                        && orbit 5    & orbit 4     && orbit 5       & orbit 6\&7 \\
\hline
\multicolumn{13}{c}{Model: \textsc{tbnew*(simpl*diskbb+constant*diskbb+gaussian)}} \\
\hline
TBNEW  &  $N_{\rm H}$ ($\times10^{22}~{\rm cm}^{-2}$) &  \multicolumn{2}{c}{$0.89 \pm 0.01$} && \multicolumn{2}{c}{$0.89 \pm 0.01$} 
                                                      && \multicolumn{2}{c}{$0.89 \pm 0.01$} && \multicolumn{2}{c}{$0.89 \pm 0.01$} \\
DISKBB &  $kT_{\rm in}$ (keV)   &  $0.693 \pm 0.004$   & $0.708 \pm 0.004$  
                                && $0.671 \pm 0.006$   & $0.694 \pm 0.005$  
                                && $0.651 \pm 0.005$   & $0.678 \pm 0.005$ 
                                && $0.640 \pm 0.004$   & $0.645 \pm 0.003$  \\
                                
       &  Norm                  &  $11624 \pm 739$    & $8234 \pm 574$   
                                && $14758 \pm 1418$   & $9480 \pm 872$ 
                                && $15153 \pm 1212$   & $8812 \pm 1036$ 
                                && $10981 \pm 796$    & $9187 \pm 495$  \\
                                
SIMPL  &  $\Gamma$              &  $3.48 \pm 0.04$   & $3.30 \pm 0.05$  
                                && $3.37 \pm 0.06$   & $3.46 \pm 0.08$  
                                && $3.46 \pm 0.05$   & $3.36 \pm 0.10$ 
                                && $3.35 \pm 0.05$   & $3.27 \pm 0.04$  \\
                                
CONSTANT &  effective $f_{\rm SC}$ &  $0.27 \pm 0.01$   & $0.20 \pm 0.01$ 
                                && $0.33 \pm 0.02$   & $0.23 \pm 0.02$  
                                && $0.33 \pm 0.02$   & $0.21 \pm 0.02$ 
                                && $0.26 \pm 0.01$   & $0.22 \pm 0.01$  \\
                                
GAUSSIAN &  $E_{\rm line}$ (keV)&  $6.4^a$           & $6.4^a$  
                                && $6.4^a$           & $6.4^a$                
                                && $6.4^a$           & $6.4^a$      
                                && $6.42 \pm 0.13$   & $6.48 \pm 0.08$  \\
                                
         &  $\sigma$ (keV)      &  $0.56 \pm 0.10$   & $0.62 \pm 0.12$  
                                && $0.5^a$           & $0.5^a$                 
                                && $0.5^a$           & $0.5^a$        
                                && $0.57 \pm 0.13$   & $0.49 \pm 0.09$  \\
                                
         &  Norm                &  $0.024 \pm 0.006$ & $0.030 \pm 0.007$  
                                && $0.017 \pm 0.006$ & $0.024 \pm 0.005$                  
                                && $0.016 \pm 0.004$ & $0.023 \pm 0.006$  
                                && $0.021 \pm 0.006$ & $0.018 \pm 0.004$  \\
\hline
         &  $\chi^{2}/{\rm dof}$ &  \multicolumn{2}{c}{422.2/447}  &&  \multicolumn{2}{c}{462.9/445 } 
                                 && \multicolumn{2}{c}{461.9/436}  &&  \multicolumn{2}{c}{457.6/445 }\\
\hline
\multicolumn{13}{l}{$^a$ Fixed.} \\
\multicolumn{13}{l}{$^b$ 0.6--10 keV unabsorbed flux in units of $10^{-8}$ erg cm$^{-2}$ s$^{-1}$.}
\end{tabular}
\end{table}
\end{landscape}

\bsp	% typesetting comment
\label{lastpage}
\end{document}